\documentclass[nologo,11pt,a4paper]{ETHpaper}

\usepackage{tabularx}                  \usepackage{MnSymbol}
\usepackage{booktabs}
\usepackage{subcaption}
\usepackage{hhline}
\usepackage[table,dvipsnames]{xcolor}
\usepackage{graphicx}                  \usepackage[english]{babel}            \usepackage{natbib}
\usepackage{setspace}
\usepackage{dcolumn}
\usepackage{setspace}

\setcitestyle{authoryear, open={(},close={)}}

\graphicspath{{figures/}}

\title{Network embeddedness indicates the innovation potential of firms}
\titlealternative{Network embeddedness indicates the innovation potential of firms}
\authoralternative{}
\author{Giacomo Vaccario, Luca Verginer, Antonios Garas, Mario V. Tomasello and Frank Schweitzer}
\authoralternative{G. Vaccario,  L. Verginer, A. Garas, M. V. Tomasello, F. Schweitzer}
\address{Chair of Systems Design, ETH Zurich, Switzerland\\
  \url{www.sg.ethz.ch}}
\www{\url{http://www.sg.ethz.ch}}

\makeframing%

\begin{document}

\maketitle%

\begin{abstract}
Firms' innovation potential depends on their position in the R\&D network. But details on this relation remain unclear because measures to quantify network embeddedness have been controversially discussed. We propose and validate a new measure, \emph{coreness}, obtained from the weighted $k$-core decomposition of the R\&D network. Using data on R\&D alliances, we analyse the change of coreness for 14,000 firms over 25 years and patenting activity. A regression analysis demonstrates that coreness explains firms' R\&D output by predicting future patenting.
\end{abstract}

\section{Introduction}\label{sec:introduction}

The ability of firms to innovate considerably depends on their research and development (R\&D) collaborations \citep{schilling2007interfirm}.
To formalize these collaborations, firms establish R\&D alliances which allow them to coordinate their activities, share resources, and exchange knowledge \citep{nooteboom1999inter}.
Alliances are publicly announced. Therefore, we know that firms engage in various alliances at the same time and change them over time~\citep{Tomasello2013}.
Data on 21,500 alliances between 14,000 firms allows us to reconstruct an alliance network, in which nodes
represent firms and links their R\&D collaborations.
This network evolves as new firms enter, incumbents exit, new alliances are formed, and established alliances end.
These processes continuously change a firm's position in the R\&D network and thus affect its ability to innovate.

This paper focuses on the relation between innovation output, as measured by the number of patents, and the importance of individual firms, as measured by their topological position in the alliance network.
It has been established in the literature that the interfirm network to which firms belong is a driver of their innovation output~\citep{freeman1991networks,powell1999network, ahuja2000collaboratio, owen2004knowledge, schilling2007interfirm, phelps2010alongitudinal}.
For example, \citet{schilling2007interfirm} have argued that global topological properties of
alliance networks (i.e., average path length and clustering coefficient) influence the number of patents.
Similarly, \citet{owen2004knowledge} have shown that firms' innovation output depends on their ability to absorb information flows in these networks.

The relation between innovation ability and network properties points to the central question of how to quantify the position of firms in the R\&D network.
A common proxy is betweenness centrality \citep{owen2004knowledge,gilsing2008network,paier2011determinants}.
It measures how often a focal node is included in the shortest paths between all existing nodes in a network \citep{newman2010introduction}.
This quantifies the importance of the focal node in controlling the information flow between other nodes and its access to information.
Other centrality measures, e.g., degree, closeness or eigenvector centrality, capture the influence
of different topological properties, such as the number of neighbours, the topological distance to all
other nodes, or the impact of neighbouring nodes.
Several of these measures have been used in various studies to determine the importance of firms and describe their \emph{embeddedness} in the alliance network \citep{powell1999network,ahuja2000collaboratio,schilling2007interfirm}.

The concept of embeddedness is pervasive in economics and sociology~\citep{polanyi1944great, granovetter1985economic}.
Yet, it is only loosely defined.
A central idea is that individuals or firms are \textit{embedded} in social relations, which in turn affect their economic behaviour.
Precisely, \citet{granovetter1985economic} stresses that ``the level of embeddedness of economic behaviour [\ldots] has always been and continues to be more substantial than is allowed for by formalists and economists''.
Also, \citet{gulati2000strategic} write, ``the image of atomistic actors competing for profits against each other in an impersonal marketplace is increasingly inadequate in a world in which firms are embedded in networks of social, professional, and exchange relationships with other organizational actors''.
Our work builds on these arguments about the importance of embeddedness.
We study how firms' innovation output is affected by their embeddedness in R\&D alliances, which we quantify in a novel method called ``coreness''.

Importantly, formal agreements such as R\&D alliances define not only economic but also social relations, as \citep{powell1999network} and \citep{gulati2000strategic} point out.
They provide access to complementary capabilities and the opportunity for learning~\citep{nooteboom1999inter}.
At the same time, establishing and maintaining social relations is costly~\citep{granovetter1985economic, uzzi1997social}.
For this reason, it is debated whether R\&D alliances have a net positive effect on firm innovation.
Indeed, when looking at the biotechnology industry, \citet{powell1999network} found that more embedded firms file more patents.
However, when considering the subsidiaries of multinational pharmaceutical companies, \citet{al2011chapter} found the opposite effect, i.e., the more a firm is embedded, the fewer patents it files.

We contribute to the ongoing discussion in two ways.
Firstly, for our analysis, we introduce a new measure of firms' embeddedness in the alliance network based on the \emph{weighted} $k$-core centrality~\citep{Garas2012a}
This is an extension of the unweighted version introduced by~\citet{seidman1983network,bollobas1989graph}.
The weighted $k$-core centrality has the advantage to control for repeated interactions, while the unweighted centrality treats repeated interactions as a single interaction.
For example, if a firm has two alliances with the same firm (i.e., repeated interactions), this
would be treated as only one alliance.
Such a simplification would underestimate a firms' involvement in R\&D collaborations.
The weighted $k$-core centrality addresses this problem using the weighted degree, which accounts for repeated interactions.

Secondly, we empirically test whether firm embeddedness has a positive or a negative effect on firms' innovation.
We operationalize innovation output as the number of patents filed by a firm.
For our analysis, we combine two databases: (i) the NBER patent database containing information on almost three million patents~\citep{hall2001nber}, and (ii) the SDC Platinum alliances database listing 21,572 alliances involving 13,936 firms between 1984 and 2009~\citep{sdc2013}.

From our analysis, we find that firms' embeddedness correlates with their innovation output and significantly affects their future innovation output.
Specifically, a regression analysis shows that firms with a higher weighted $k$-core centrality in a given year also have a higher innovation output in the following year.
The results are robust and significant even after controlling for the number of previously filed patents, industrial sector, and other measures capturing firms position in the alliance network.

The remaining of this paper is divided as follows.
We first present the data to infer R\&D alliances and measure innovation output via patent data.
We then introduce the $k$-core centrality to operationalize our notion of embeddedness.
We describe the regression analysis used to show the effect of a firm's embeddedness on its patenting activity.
In the Results section, we illustrate the evolution of the alliance network and offer results supporting the hypothesis that embeddedness is indeed correlated with patenting activity.
Further evidence in support of this result is then provided by showing the results of the regression analysis.
Finally, in the discussion, we summarize the findings and discuss how they fit in the extant literature.

\section{Materials and methods}\label{SEC:Data}
\subsection{Data sets}\label{sec:data-sets}

In this paper, we build on two data sets.
The first data set, obtained from Thomson Reuters' SDC Platinum alliances database, contains
all publicly announced R\&D partnerships between firms~\citep{sdc2013}.
Because the SDC database does not provide a unique identifier for each firm, we use the firm names reported in the dataset.
Therefore, we disambiguate names, i.e., we correct for the cases where two or more entries with different names corresponded to the same firm, by manually controlling for spelling, legal extensions (e.g., LTD, INC), and any other recurrent keywords (e.g., BIO, TECH, PHARMA, LAB).
We keep subsidiaries of a firm located in different countries as separated entities.

In total, we use information about 21,572 alliances involving 13,936 firms between 1984 and 2009.
We note that these alliances can involve different economic actors, e.g., universities, but we refer to them as firms.
We used a firm's 4-digit Standard Industrial Classification (SIC) code to classify the industrial sector.
The data allows us to reconstruct a network in which
nodes represent firms, and links represent R\&D collaborations.
On this network, we perform a weighted $k$-core decomposition to compute the coreness value of each
firm as a proxy of its embeddedness in the R\&D network.

The second data set is obtained from the NBER patent database of the National Bureau of Economic Research \citep{hall2001nber}.
It contains detailed information on almost three million patents granted in the U.S.A. between 1974 and 2006.
Every patent is assigned to one or more assignees and is classified according to the International Patent Classification (IPC) system.
We use this data to estimate the innovation output of each firm by means of its number of patents in the respective time window.

\subsection{Reconstructing the R\&D network}\label{sec:reconstr-cumm-netw}

Because the network of firm collaborations is highly dynamic, there are various ways of studying its evolution over time.
Work by \citet{Tomasello2013} focuses on the changing growth pattern in consecutive 5-year time intervals to reveal a remarkable life cycle dynamics of the R\&D network over the whole period of 25 years.
Here, we focus on how a firm's (current) position, i.e., embeddedness, in the network affects its innovation output.

\begin{figure}[htbp]
  \centering
  \includegraphics[width=0.245\textwidth]{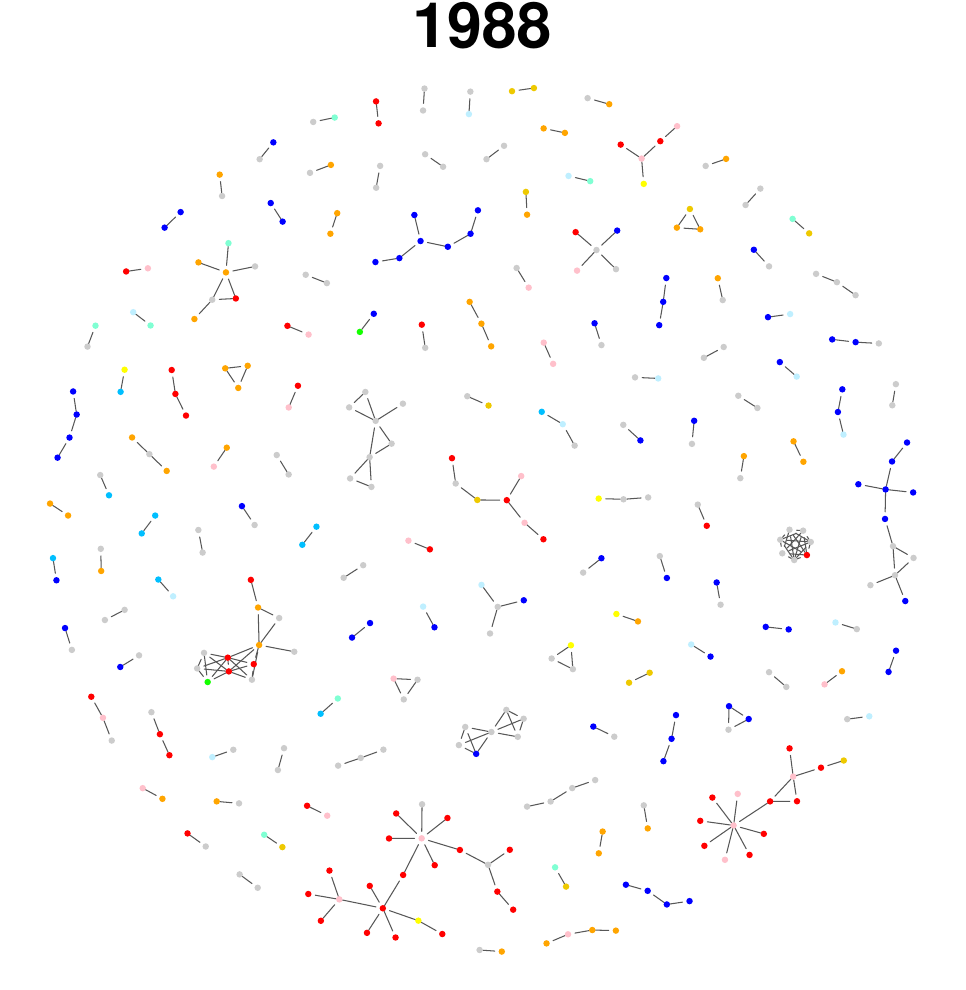}\hfil
  \includegraphics[width=0.245\textwidth]{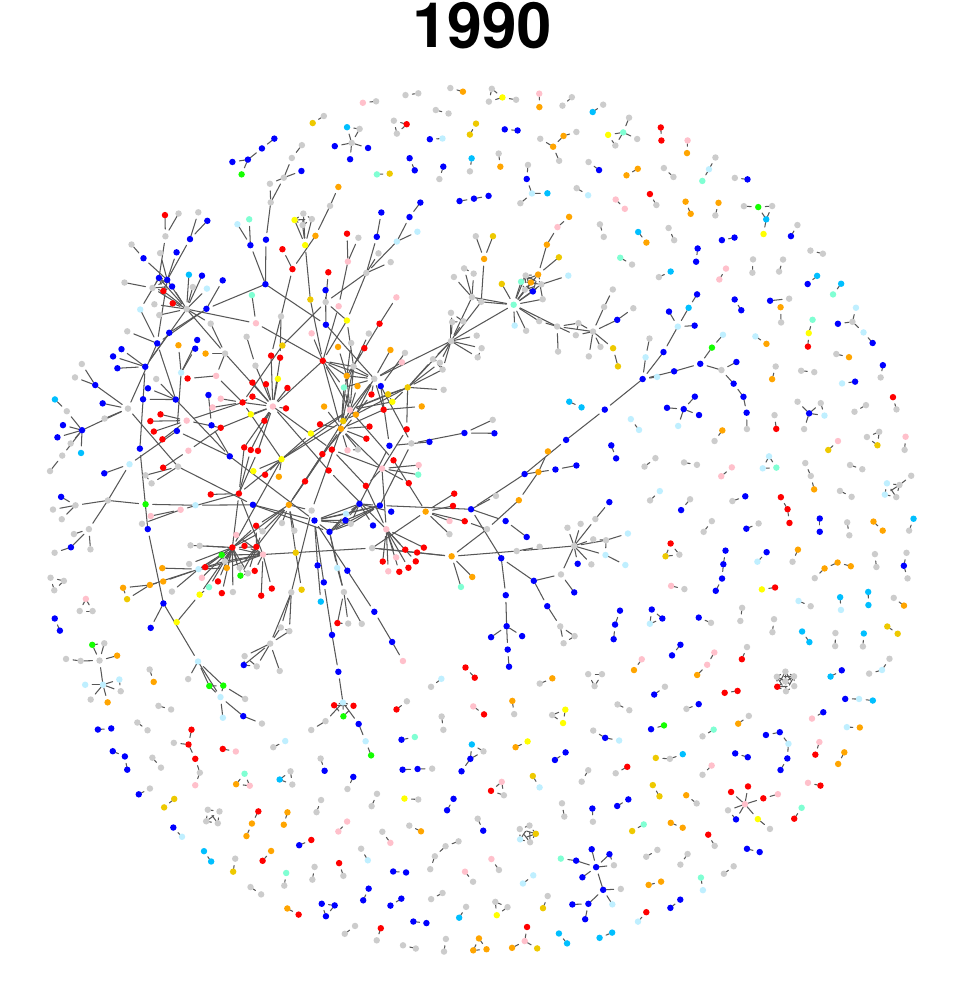}\hfil
  \includegraphics[width=0.245\textwidth]{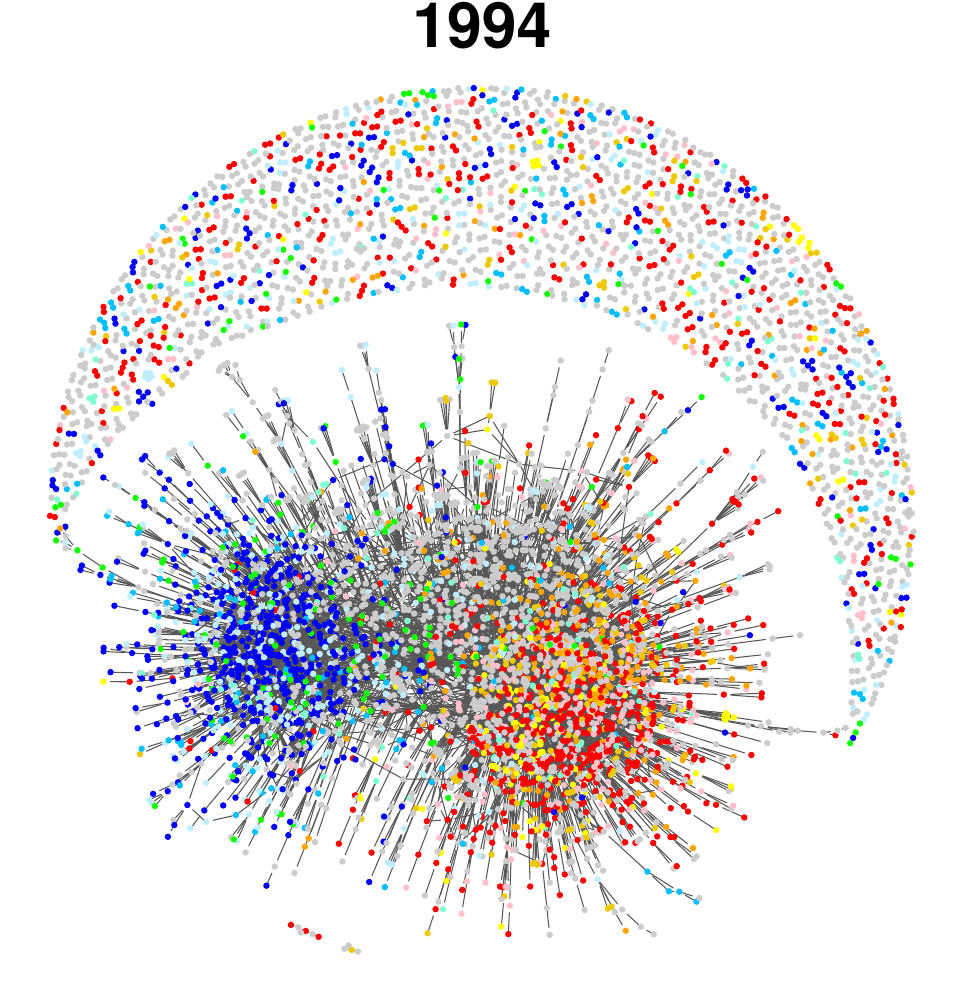}\hfil
  \includegraphics[width=0.245\textwidth]{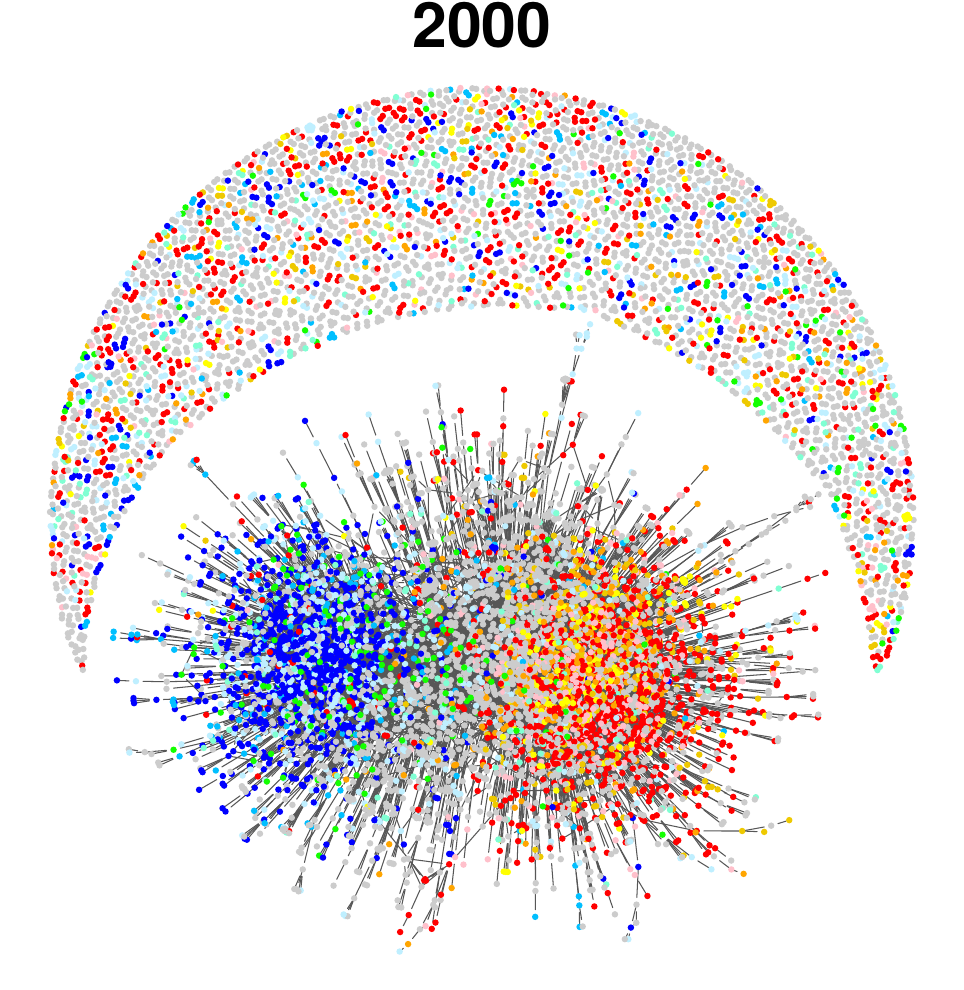}\\
  \includegraphics[width=0.4\textwidth]{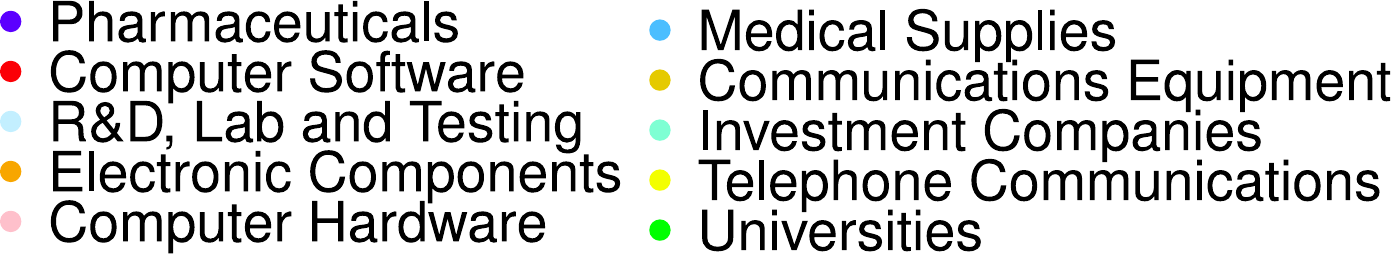}\\
  \caption{Snapshots of the R\&D network showing its evolution and the emergence of a large connected component.}\label{fig:cum}
\end{figure}

We start by reconstructing the cumulative R\&D network: we use as time resolution one year and add a new link to the cumulative network every time an alliance of two firms is listed in the dataset in this time window.
When an alliance involves more than two firms (consortium), all the firms involved are connected in pairs, resulting in a fully connected clique.
The weight $w^{ij}(t)$ of a link indicates the total number of alliances between firms $i$ and $j$ up to time $t$.
If, during the same time interval, two firms $i$ and $j$ have more than one collaboration on different projects, such multiple links are also considered in the weight.

\subsection{Quantifying network embeddedness}\label{SEC:Methods}

From the reconstructed cumulative R\&D network, we can compute the embeddedness of nodes in this network.
The measure used to proxy embeddedness is based on \emph{coreness} $C_{C}^{i}$.
For a given network, a coreness value can be assigned to every node using the $k$-core decomposition.
This procedure recursively removes all nodes with a degree less than $d$ until only nodes with a degree equal to or larger than $d$ remain.
The procedure starts with $d=1$, i.e., it removes all nodes that have only one neighbour in the networks.
The removal may leave these neighbouring nodes with a single neighbour.
In the second step of the procedure, such nodes are also removed.
Their removal again may leave other nodes with one remaining neighbour.
Thus, in the third step, they are also removed and so forth, unless no nodes can be removed.
Then, all nodes that have been removed during this procedure are assigned a shell number $k_{s}$ equal to $d$.

Nodes with a small values of $k_{s}$ are removed very early because they are topologically weakly embedded in the network.
Conversely, nodes with the largest value of $k_{s}=k_{s}^{\max}$ form the core of the network.
The difference between the $k^{i}_{s}$ value of a node $i$ and the value of the core is called \emph{coreness}, or distance from the core, $C_{C}^{i}=k_{s}^{\max}-k_{s}^{i}$.
Nodes close to the core, i.e. with \emph{small} coreness values, are well embedded in the network,
while nodes with \emph{large} coreness values are not.

Since the method described uses only information about the node degree and ignores the link weights, it is called unweighted $k$-core decomposition.
This method has been successfully applied to characterize various real-world networks~\citep{Carmi2007,Garas2010}.
Moreover, \citet{Kitsak2010} showed that the coreness value of a node is a more accurate predictor of its spreading potential than, for example, its degree.

This paper uses an extension of the unweighted $k$-core decomposition, which also considers the link weights.
The weighted $k$-core decomposition~\citep{Garas2012a} uses the same procedure to remove nodes as the unweighted version, but a refined measure for the node degree, called weighted degree, $d'$.
The weighted degree, $d'$, depends on two free parameters $\alpha$ and $\beta$ balancing the influence of the weights $w_{ij}$ which indicate multiple alliances between the same firms.
Similar to~\citet{Garas2012a}, we consider the case when $\alpha=\beta=1$.
With this choice, we assign to the weight and the degree the same importance, and the equation for the weighted degree of node $i$ becomes a geometric mean:

\begin{equation}\label{eq:gsh.kprime}
    d'_{i}={\left(d_{i}^\alpha {\left( \sum\nolimits_{j}^{d_{i}}{w_{ij}} \right)}^\beta \right)}^{\frac{1}{\alpha+\beta}} \xrightarrow[\alpha=\beta=1]{} \quad
    d'_{i}=\sqrt{d_{i} \left( \sum\nolimits_{j}^{d_{i}}{w_{ij}} \right)}
\end{equation}

$d^{i}$ is the degree of node $i$ and $w^{ij}$ is the weight of the link between nodes $i$ and $j$.
The summation goes over all neighbours of $i$.

Note that the coreness value in a dynamic network, such as the alliance network, changes over time.
Specifically, there are two processes affecting the coreness of a firm $i$: (i) formation of alliances involving $i$ (direct) and (ii) formation of alliances which not involving $i$ (indirect).

\begin{figure}[htbp]
    \centering
      \includegraphics[width=0.48\textwidth]{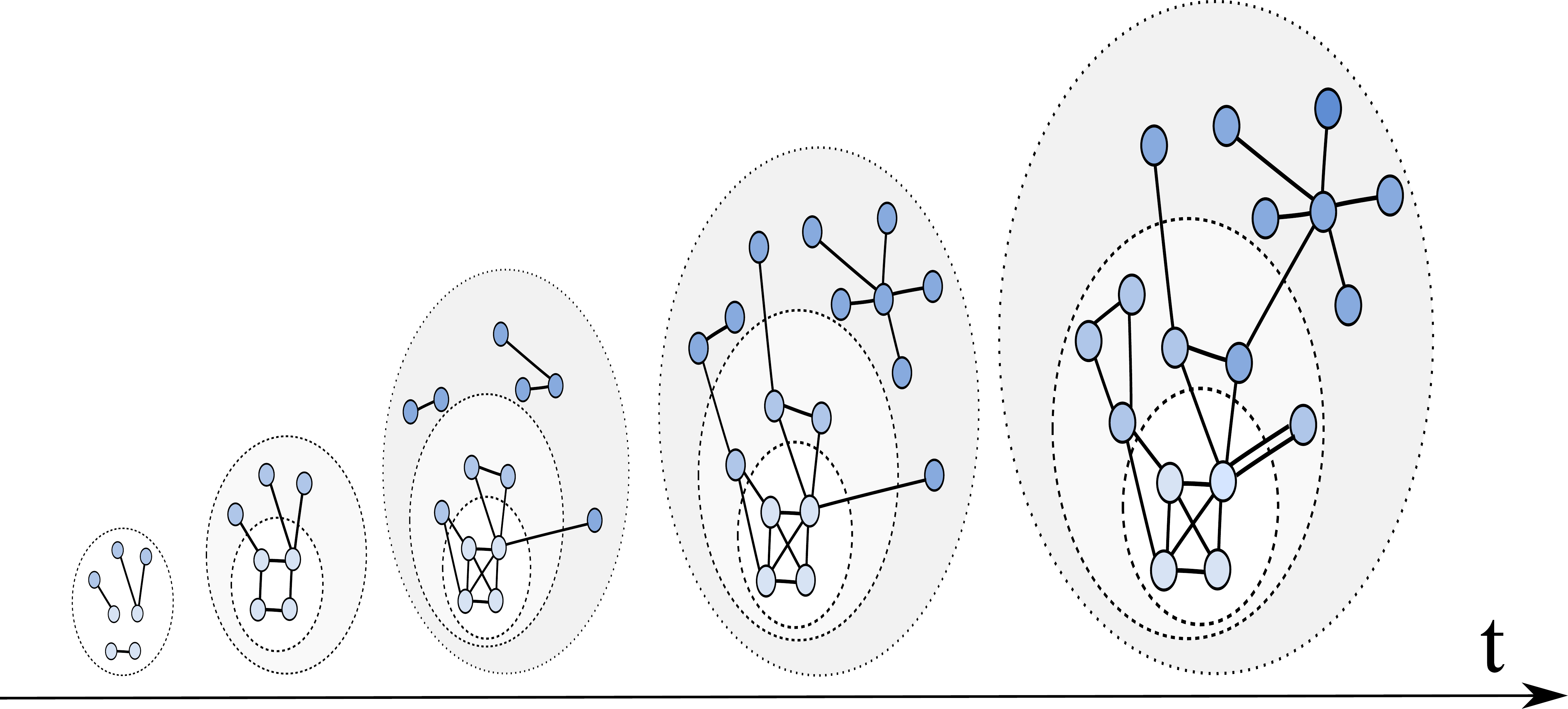}
    \caption{Illustration of the network evolution where new $k$-shells emerge as new links are formed.}\label{fig:kcoresketch}
  \end{figure}

  In Figure~\ref{fig:kcoresketch} we provide a simplified example of a growing network where similar to real R\&D networks, new firms enter the network by creating new links either with existing firms or newcomers.
  In the first process firm, $i$ plays an active role in forming new alliances, and its coreness may change accordingly.
  In the second process, firm $i$ plays no direct role in forming new alliances. However, the network grows, and new shells emerge.
  Therefore, a firm $i$'s position may still change.
  This implies that the coreness of a particular firm $i$ may change even without any new R\&D alliances involving that firm.

  All coreness values reported in this paper are based on the \emph{weighted} $k$-core decomposition.
  This differs from the work by~\citet{powell1999network}, where the Katz-Bonacich centrality is used to operationalize embeddedness.
  This centrality is similar to the known eigenvector centrality, i.e. it considers the weight of direct and indirect neighbours in measuring the importance of nodes.
  We argue that the Katz-Bonacich centrality has substantial limitations in measuring embeddedness because it may reflect the degree of a node, which is not embeddedness.

\begin{figure}
  \centering
  \includegraphics[width=0.65\textwidth]{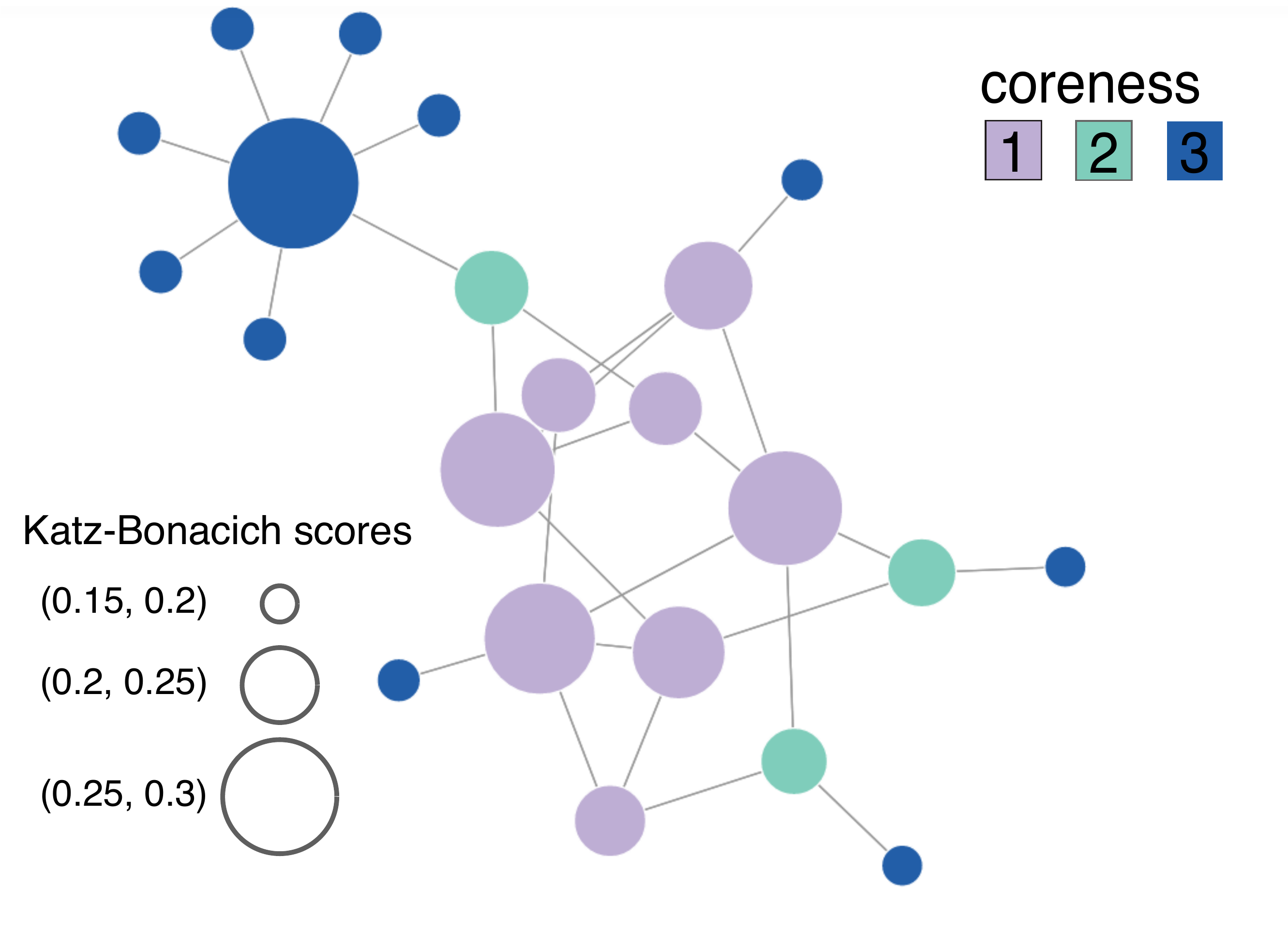}
  \caption{Differences between degree, Katz-Bonacich centrality, and coreness. We illustrate a
    undirected network where the Katz-Bonacich centrality fails to capture the embeddedness of nodes
    in a network as it mainly captures degree.}\label{fig:kats-vs-K-core}
\end{figure}

To illustrate this argument, in Figure~\ref{fig:kats-vs-K-core} we plot a network where nodes are characterized by three different common measures.
The node's size indicates its Katz-Bonacich centrality, the colour of a node indicates its coreness value, and the number of links of a node indicates its degree-centrality.
We see that nodes with a high degree also have a large size, i.e., a higher Katz-Bonacich centrality.
However, these nodes are not necessarily well embedded as the node in the upper left part of Figure~\ref{fig:kats-vs-K-core} shows.
Despite its high degree and high Katz-Bonacich centrality, this node can be easily disconnected from the network by removing one link; hence it is not well embedded.
The coreness measure addresses this problem by assigning a low coreness value to nodes that remain connected after the link removal procedure.

\subsection{Regression variables}\label{sec:data-aggregation}

For the regression analysis, we follow the approach of~\citet{schilling2007interfirm} who reconstructed the alliance network by aggregating alliances in time windows of three years.
This aggregation yields an unbalanced panel because a firm can ``disappear'' if it does not form alliances in all time windows.
Nevertheless, alternative approaches would result in other problems.
For example, considering time-aggregated alliance networks with increasing time windows as done in~\citep{tomasello2014therole,vaccario2018quantifying} would have the drawback that new alliances have the same weight as much older alliances.
This assumption is unreasonable to study how alliances affect innovation output.
Empirical studies have shown that alliances end after about three years~\citep{phelps2010alongitudinal}.
Therefore, in line with \citet{schilling2007interfirm}, we consider a time window of three years.

To work with the time windows, one could carry out a regression for each year separately and include only active firms.
However, by adopting this approach would hinder the comparison of parameters across models.
Alternatively, one could restrict the analysis to years with more firms forming alliances and filing patents.
This approach, however, creates a bias because outcome variables would be used to prepare the data sample for the regression.
For these reasons, we opt for a pool regression and use all firms' observations from different years.
Moreover, we control for (i) sector and (ii) time fixed effects.

The \emph{dependent} variable is the innovation output of a firm, which we quantify using the number of patents filed by a firm in the next year.
The \emph{independent} variable is the firms' embeddedness, which we quantify using the coreness value $C$.
For each firm, we control for
the number of patents published in the previous five years, the
number of partners,
betweenness centrality,
local efficiency,
local clustering coefficient, and
local reach for the following reasons.

We control for the number of partners because \citet{shan1994interfirm} have shown that innovation output is significantly affected by the number of cooperative relationships.
Moreover, \citet{nohria1992networks} found that interfirm cooperation rises with their size, and hence, the number of partners is also a control for the firm size.
\citet{owen2004knowledge} proxied a firm's ability to absorb information flows via betweenness centrality and shown that it affects firms' innovation.
The local clustering coefficient measures how many neighbours of a focal node are also connected, this way forming cliques or clusters.
Local reach quantifies the fraction of all nodes in a network reachable from a focal node.
Both measures are the local versions of the variables investigated by~\citep{schilling2007interfirm} which argued for their importance.
Also, they found that local efficiency, i.e., to which extent a firm's partners are non-redundant, has a significant effect.
Finally, we control for industrial sectors as different sectors have different patenting practices.

In Table~\ref{tab:variables}, we summarize the variables that we will use to model firms' innovation output.
\citet{schilling2007interfirm} and \citet{owen2004knowledge} discuss these variables and their economic meaning exhaustively. The interested reader can refer to them.

\begin{table}
  \footnotesize
  \centering
  \begin{tabular}{rc}
  \toprule
  \textbf{Variables} &\\
  \textbf{Dependent}       & Number of patents between $t$ and $t+1$, $P_i(t+1)$\\
  \textbf{Independent} &  Coreness $C_i(t)$\\
  \midrule
  \textbf{Control} & \\
  \textbf{Firm-level}                       &                                      \\
  Pre-sample of patents                       & \# of patents in $[t-5, t]$        \\
  Number of partners & $d_i(t)$ \\
  Betweenness centrality                       & $b_i(t)$\\
  Local clustering coefficient & $2e_i/(d_i(d_i - 1))$\\
  Local reach & $\sum_{j\neq i}1/d(i,j)$ \\
  Local Efficiency & $\sum_j \left(1 - \sum_k \frac{A_{ik}}{\sum_l {A_{il}}}\delta_{jk} \right)$ \\
  \textbf{Network-level} & \\
  Industrial sector & a dummy for each sector\\
  \textbf{Time-level} & \\
  Year of the alliance network & a dummy for each year\\
  \midrule
  \textbf{Model specification}            &    Zero-inflated Negative Binomial            \\
  \bottomrule
  \end{tabular}
  \caption{The four types of variables used to explain firms' innovation output.
  For the details on how to compute them see \citet{schilling2007interfirm, newman2018networks, burt2009structural}.}\label{tab:variables}
\end{table}

\subsection{Regression model}\label{sec:setup-our-regression}

We use a zero-inflated negative binomial model.
The negative binomial is a good model for firms' number of patents as this number is an over-dispersed count variable.
Indeed, in one year, a single firm files between $0$ to more than $2\,000$ patents.
The mean patent count is around $17.3$, and the standard deviation is $111.24$, i.e., almost 10 times bigger than the mean.
However, we also have that more than 50\% of firms file 0 patents in a year.
We use a zero-inflated model to account for the excessive zero counts in the data.
We assume that there are two processes driving firms' innovation output.
Firms either have the capability to file patents, or they do not.
If they do not have the ability, their patent count is 0, while in the other case, we model their patent count using a negative binomial.
Note that by using a zero-inflated model, we assume that firms with the ability to file patents can still fail and then obtain zero as patent count.

We consider three different model specifications.
In all models, the independent variables and controls are computed before time $t$, while the dependent variables are computed at $t+1$.
Hence, the general form of the models is:

\begin{equation}
  P_i(t+1) = \mathcal{F}\left[C_i(t),F_i(t), I_i(t)\right]
\end{equation}

where $P^i(t+1)$ is number of patents filed by firm $i$ between $t$ and $t+1$, $C_i(t)$ is the coreness of firm $i$ at time $t$, and $F^i(t)$  and $I_i(t)$ are the controls at the firm and the network level.
\textit{Model 1} contains only the control variables used in~\citep{schilling2007interfirm}.
\textit{Model 2} contains all the variables of Model 1, and we add the local reach and local clustering coefficient, i.e., the local versions of the variables studied in~\citep{schilling2007interfirm}.
In addition, we add the number of partners as a control for firm size.
These two models are baseline models showing how well we can capture the innovation output of firms \textit{without} using the firm embeddedness.
\textit{Model 3} contains all the variables of Model 2, and we introduce the measure of firms' embeddedness, i.e., their coreness $C_i$.
For the regression models, we name our measure CORE.

\section{Results}\label{sec:results}

\subsection{Core-periphery structure}

\begin{figure}[ht]
    \centering
    \includegraphics[width=0.40\textwidth]{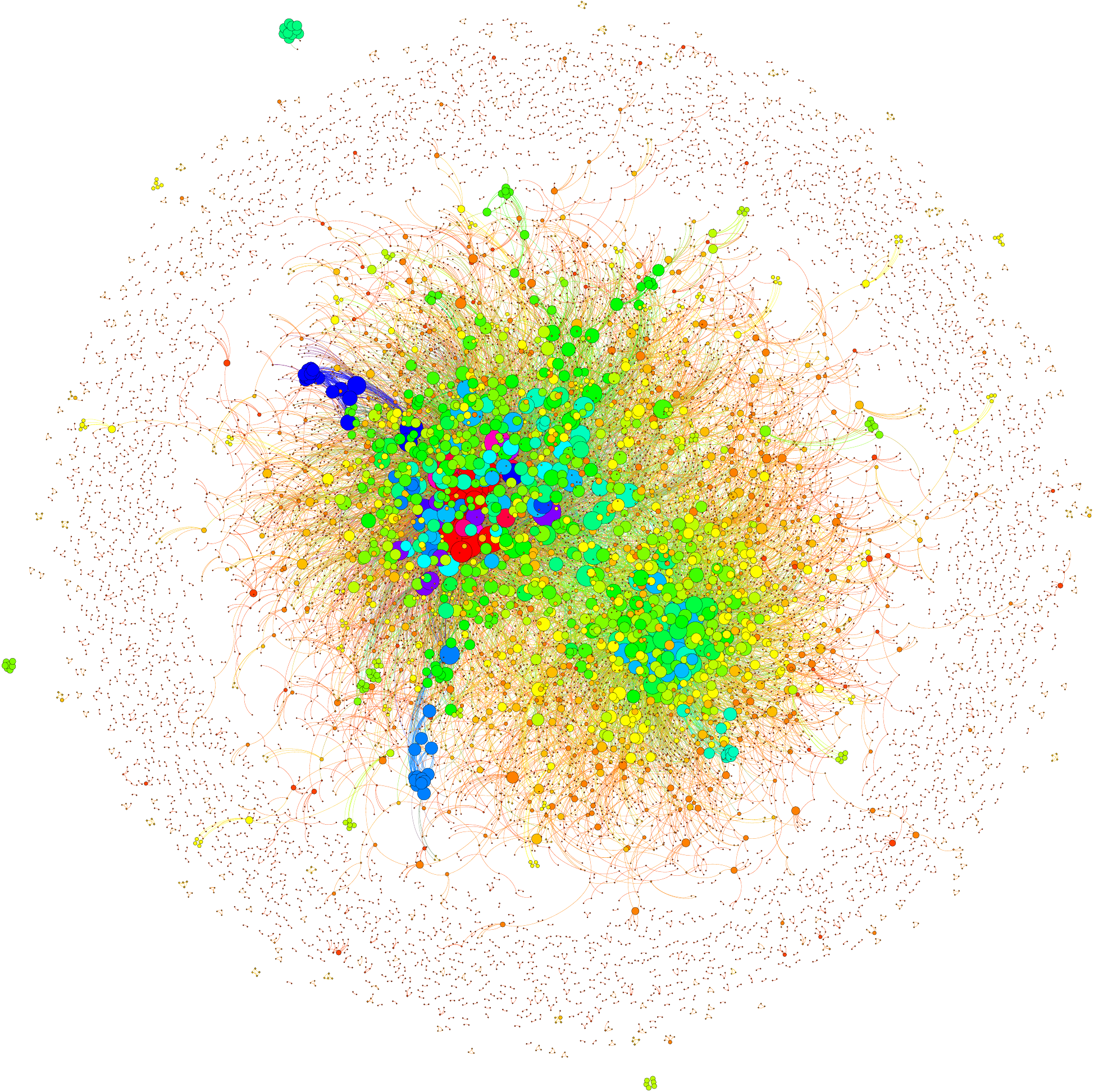}\hfill
    \includegraphics[width=0.58\textwidth]{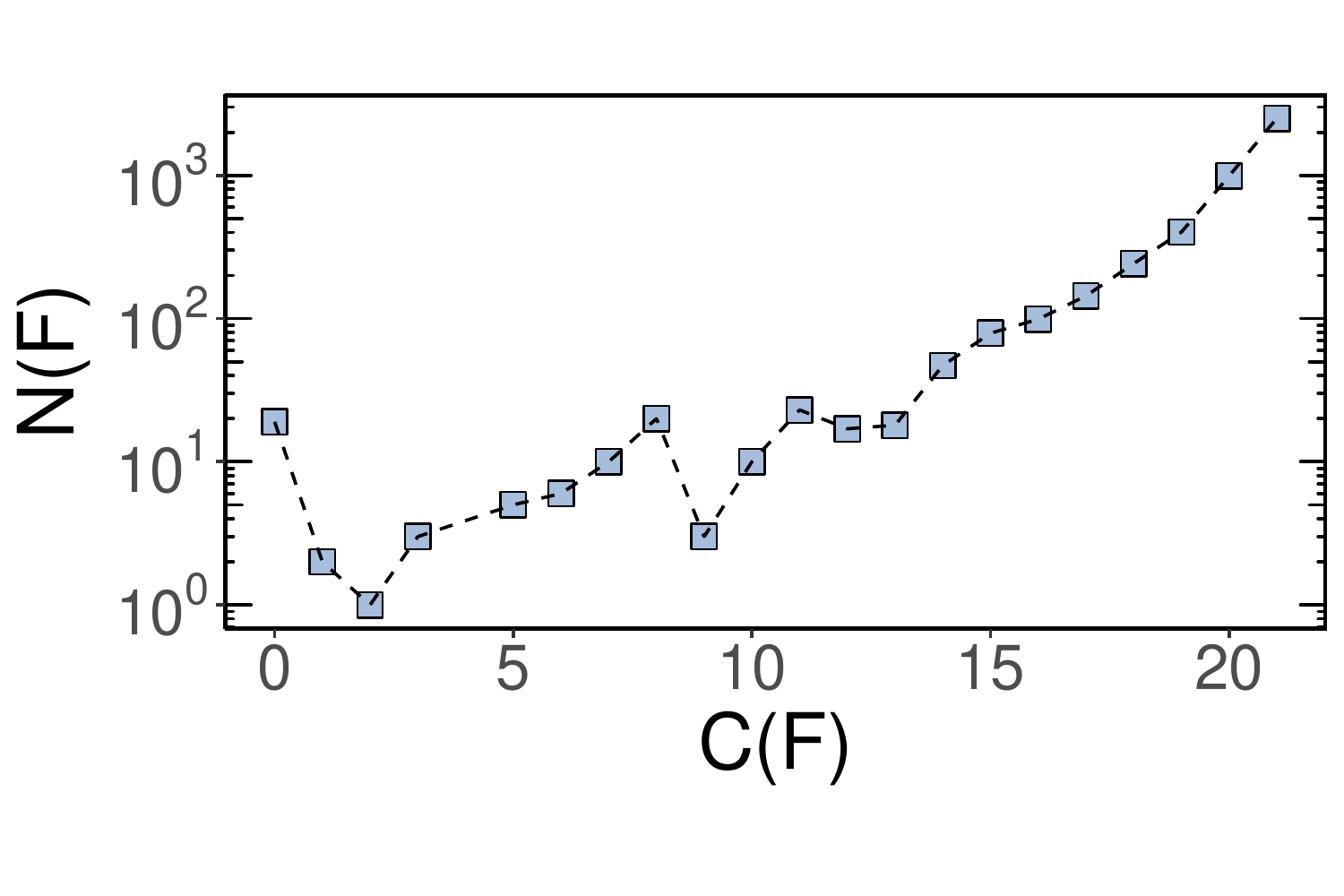}
    \caption{ \emph{(left)} Graphical representation of the cumulative R\&D network at the end of 2009.
    This plot is made with \citet{gephi} using the OpenOrd layout.
    The different colors represent different coreness values, with red assigned to the core nodes.
    \emph{(right)} The network has a strong core-periphery structure~\citep{Borgatti2000}, i.e. only a small number of nodes having small $C$ values, while the majority of the nodes are located in the periphery and have large $C$ values.}\label{fig:GTS-1}
\end{figure}

We apply the weighted $k$-core decomposition to the cumulative R\&D network to assign a coreness value to each firm.
When the network evolves over time, a firms' position in the network may change.
We take the cumulative R\&D network at the very last year, which is the year 2009, and indicate the firms' coreness value in this maximal network as $C_{i}(F)$ (F stands for final).

Figure~\ref{fig:GTS-1} (left) shows a network plot of the cumulative R\&D network in 2009.
The nodes are coloured according to their coreness value $C_{i}(F)$, and their size is proportional to their cumulative degree $d_{i}$, i.e., the cumulative number of allied partners.
The figure highlights that several firms, despite their many alliances, are not part of the core but of the periphery.

Figure~\ref{fig:GTS-1} (right) shows the frequency of coreness values, ${N}_{C(F)}$, for the cumulative R\&D network in 2009.
Taking into account that $N=13,936$, we see that the total number of firms with small coreness values, $0\leq C(F)\leq 5$, i.e. firms that are part of, or very close to, the core are relatively small, but there is a large number of firms in the periphery, $C(F)>5$.
This topological property indicates a very pronounced core-periphery structure, and hence the R\&D network exhibits a large variation of coreness values.
Given the broad distribution of coreness values, in our analysis, we can explore heterogeneous firm embeddedness.

\subsection{Correlations between network position and innovation output}\label{sec:link-centr-succ}

\begin{figure}[htbp]
  \centering
  \includegraphics[width=0.7\textwidth]{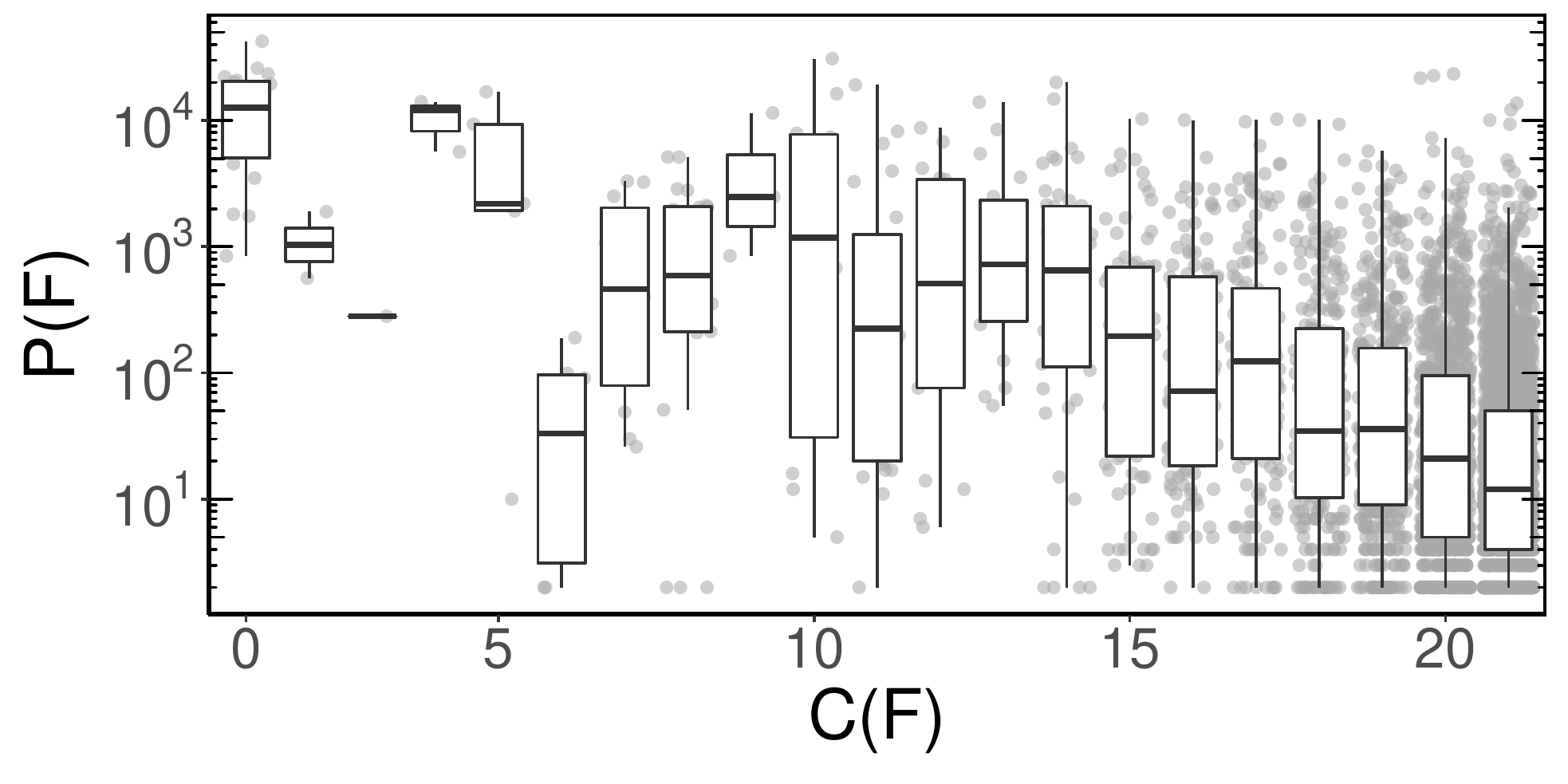}
  \caption{Box plot of the number of patents $P$ against, coreness $C(F)$. The black middle line represents the median. The top and bottom of hinges denote the 25 and 75 percentiles, respectively. The whiskers represent the 95\% confidence interval.
  The scatterplot shows the actual data points.}\label{fig:success}
\end{figure}

We explore to what extent the position of firms in the R\&D network is indicative of innovation output.
As explained in Section~\ref{sec:data-sets}, we measure the firms' innovation output via the number of patents they file.
We first show how this number correlates with the firms' coreness values.
The results are shown in Figure~\ref{fig:success}.
According to the distribution of coreness values shown in Figure~\ref{fig:GTS-1}~(right), each firm belongs to one out of 21 coreness classes indicated by $C(F)$.
In Figure~\ref{fig:success} we show the average number of patents for each of these classes.
Although the patent data is somewhat scattered, there is a clear indication that the number of patents \emph{increases} with a better network position, i.e., with smaller $C(F)$ where $C(F)=0$ indicates the core.

More precisely, by defining with $\bar{P}(F)$ the average number of patents filed by firms with coreness $C(F)$, the pearson correlation between $\bar{P}(F)$ and $C(F)$ is $-0.581$.
This means that the weighted coreness of a firm --- i.e. a topological measure --- becomes a very strong indicator of its performance in R\&D activities, as measured by the number of patents.

\begin{figure}[htbp]
  \centering
\includegraphics[width=0.48\textwidth]{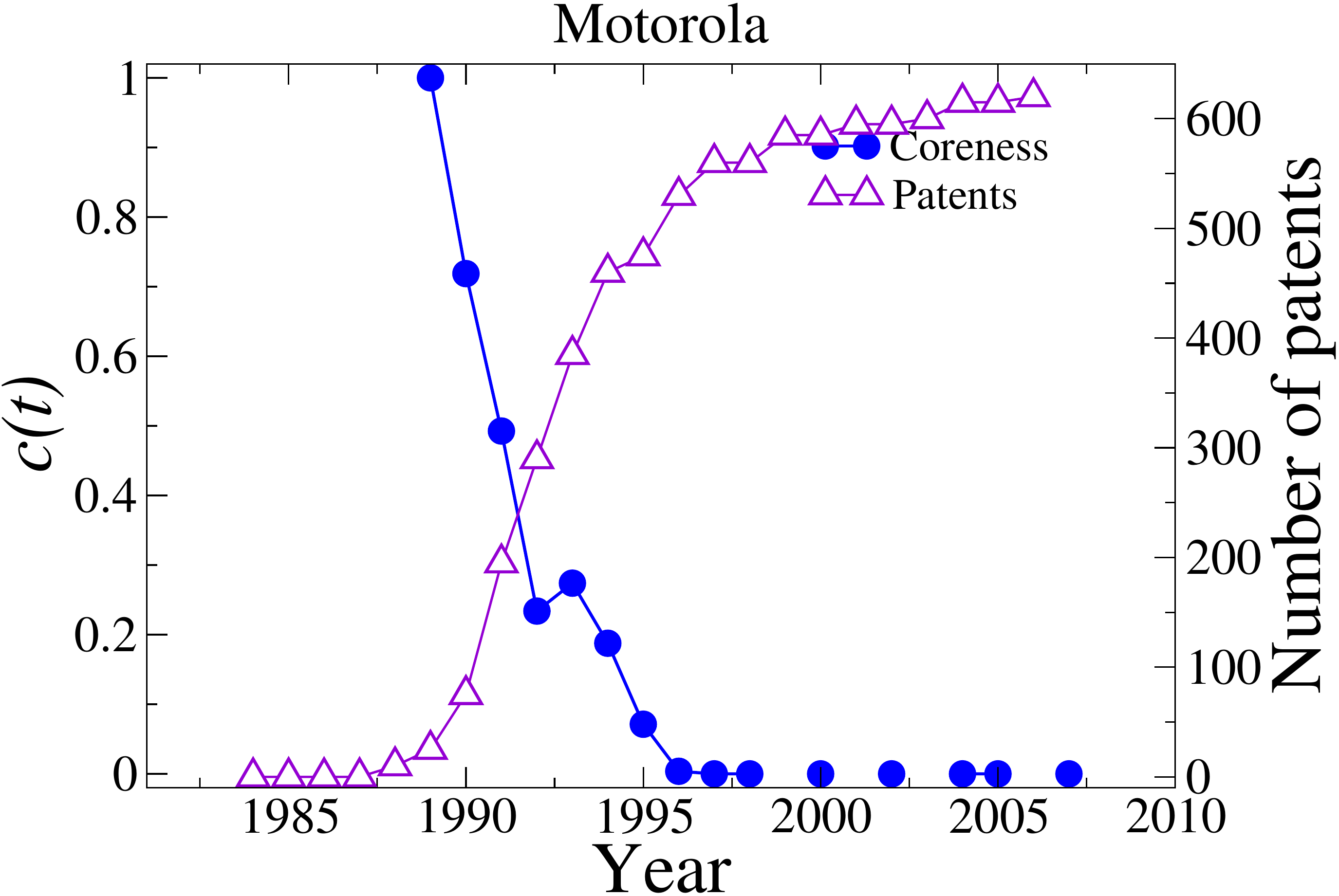} \hfill
\includegraphics[width=0.48\textwidth]{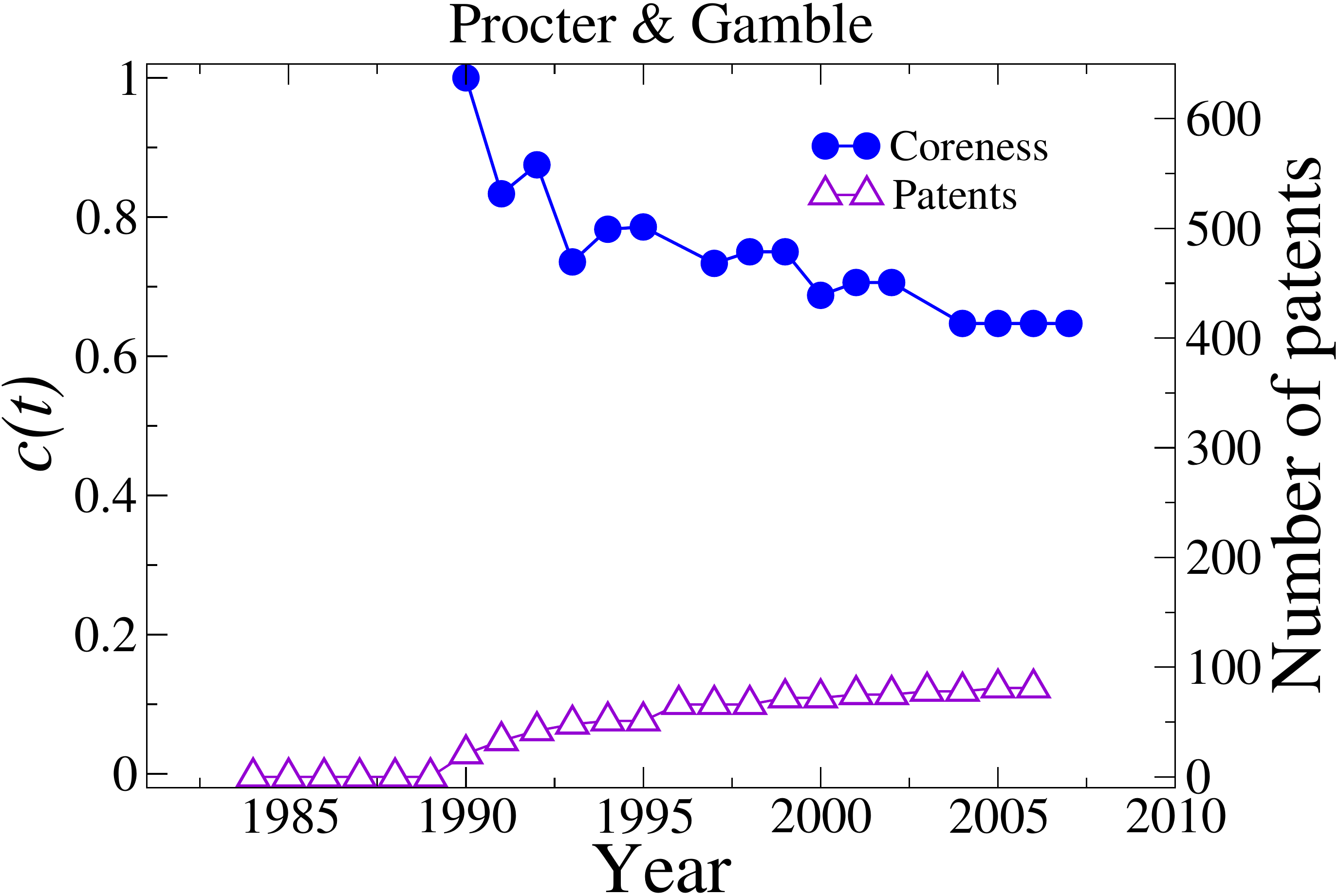} \\
\mbox{}\\
\includegraphics[width=0.48\textwidth]{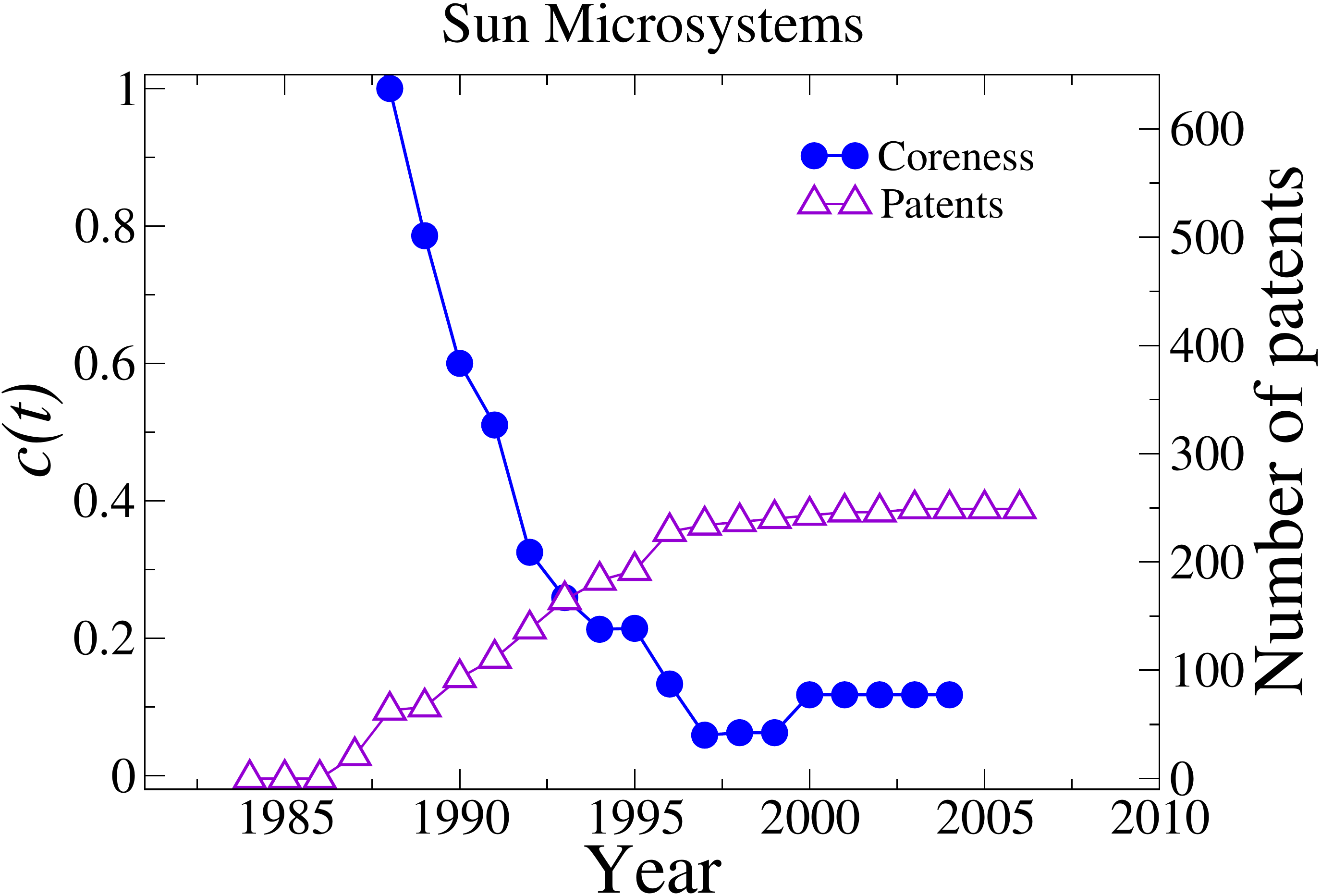} \hfill
\includegraphics[width=0.48\textwidth]{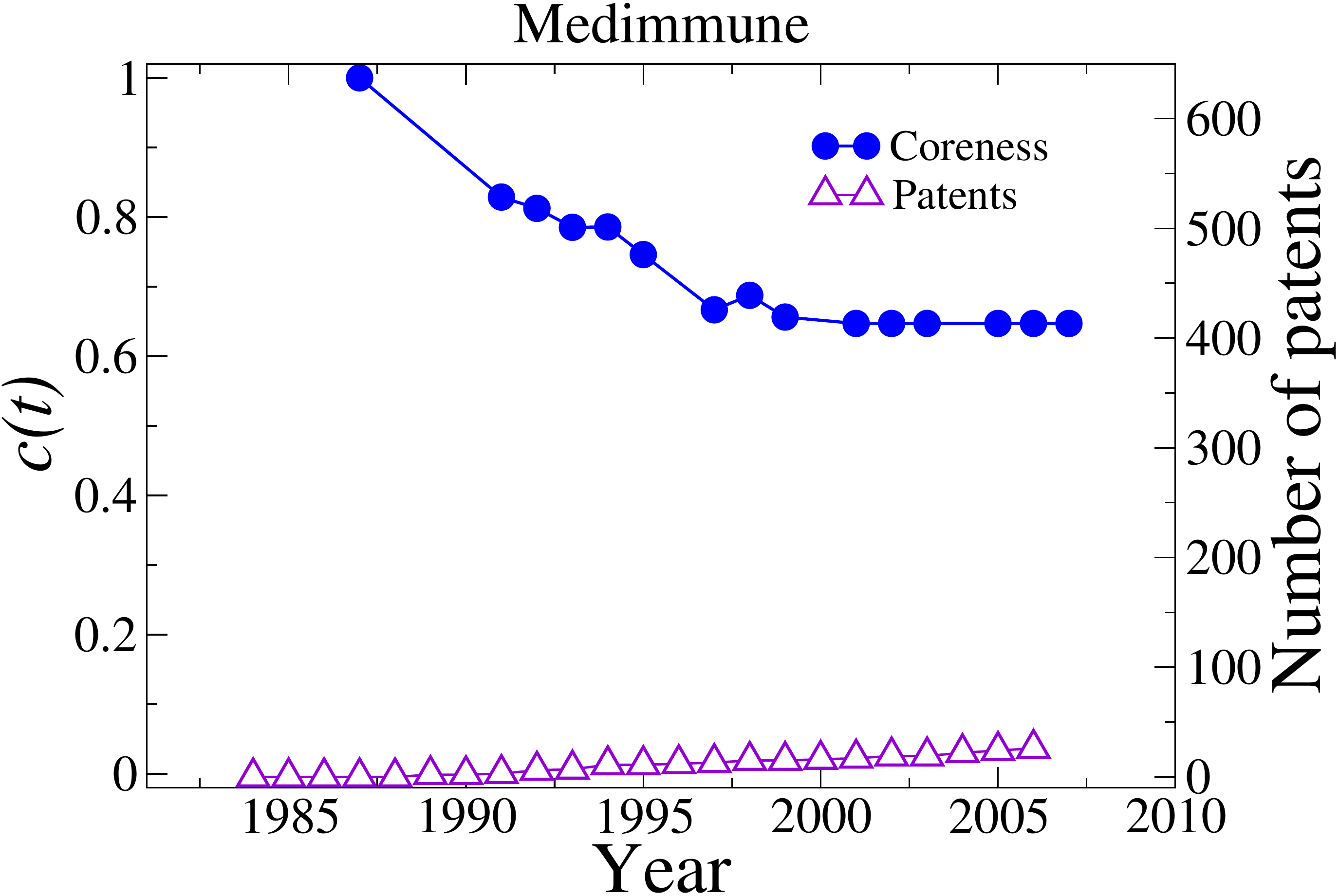}
  \caption{Coreness change and number of patents:
  (top-left) Motorola,
  (bottom-left) Sun Microsystems,
  (top-right) Procter \& Gamble,
  and (bottom-right) Medimmune.}\label{fig:cp}
\end{figure}

After showing that the coreness of a firm correlates positively with innovation output in the cumulative network, we look at its evolution over time.
To make coreness values comparable at different times, we define relative coreness  as $c_{i}(t)=C_{i}(t)/C_\text{max}(t)$, i.e. as the ratio between the current coreness $C_{i}(t)$ and the maximum coreness $C_\text{max}(t)$ at the same time
The variable $c_i(t)$ lies between 0 and 1 where 0 is the very \emph{core} and 1 to the outest \emph{periphery}.

In Figure~\ref{fig:cp}, we present two firms, \texttt{Motorola} and \texttt{Sun} that reached the core of the R\&D network and two that did not: \texttt{Procter\&Gamble} and \texttt{Medimmune}.
All four firms start with high relative coreness values.
\texttt{Motorola} and \texttt{Sun Microsystems} have a declining relative coreness which reach small values.
In other words, these two firms move closer to the core thanks to the actively formed alliances and the general evolution of the network.
This dynamic distinguishes them from \texttt{Procter\&Gamble} and \texttt{Medimmune} that remain in the periphery.
Looking at the number of patents filed, we see that the two firms moving closer to the core also file more patents, while the two firms in the periphery filed fewer.
A more detailed investigation of the coreness evolution and the possible mechanisms to reach the core are discussed in \citet{schweitzer-garas-2021}.

\subsection{Regression results}\label{sec:career-path-firms}

In Table~\ref{tab:models}, we report a summary of the regressions performed.
The first result is that the past number of filed patents, $\log(\mathrm{PAT} + 1)$, is an important control.
Precisely, this variable controls for firms' heterogeneity~\citep{blundell1995dynamic}, and hence, for their ability to either file or not file patents.
In other words, a firm's momentum in filing patents positively affects its innovation output.
This result is in line with the one obtained by \citet{schilling2007interfirm}.

Our hypothesis is that coreness has a negative effect on firms' innovation output.
This hypothesis is supported by the significant and negative effect of \texttt{CORE} in \textit{Model 3} (see Table~\ref{tab:models}).
If we compare \textit{Model 2} and \textit{Model 3}, we find that the two control variables change their significance:
The effect size of \texttt{LOCAL\_REACH} becomes indistinguishable from zero, while \texttt{EFF}
(local efficiency) becomes statistically significant.
Their change in significance means that these variables have some correlations with \texttt{CORE}.
Such correlations are partially expected because all these centrality measures tend to be correlated~\citep{Freeman1979}.
To check that these correlations do not affect our results, we performed a robustness analysis.

We create a fourth model, \textit{Model 4}, in which we remove \texttt{EFF},
the local efficiency, but keep \texttt{CORE}.
This way, we check if the effect of \texttt{CORE} is driven by the potential correlation with this variable.
For this model, the parameter of \texttt{CORE} loses about 10\% of its value (see Table~\ref{tab:models}).
However, it stays negative and significant.

\begin{figure}
  \centering
  \includegraphics{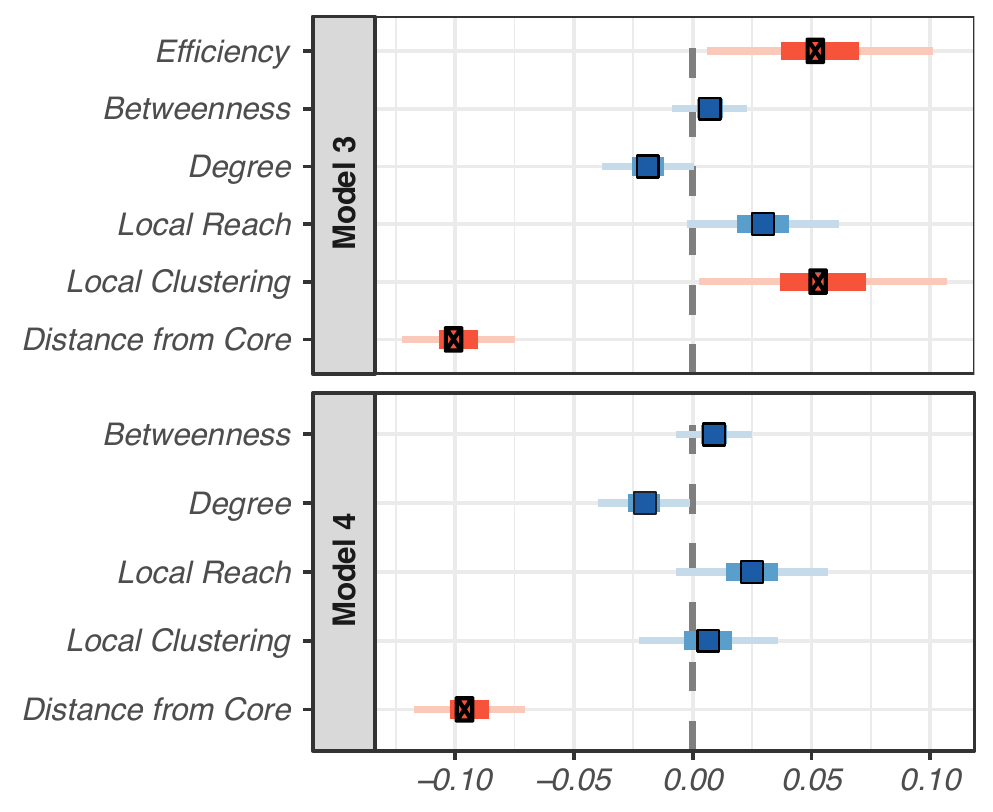}
  \caption{Coefficient plots for \textit{Model 3} and \textit{Model 4}. In blue we plot the effect
    size that are not indistinguishable from zero, while in red we plot the statistically
    significant effects. Note that in \textit{Model 4} we remove the (local) efficiency \texttt{EFF}
    and find that the effect size for the coreness \texttt{CORE} remains almost unchanged.}\label{fig:coefplot}
\end{figure}

In Figure~\ref{fig:coefplot}, we also illustrate this change by plotting the effect size of the analyzed network measures for \textit{Model 3} and \textit{Model 4}.
From this visualization, we see that the effect size of  \texttt{CORE} remains almost unchanged.
This indicates that the correlation with \texttt{EFF} increases the importance of \texttt{CORE} but does not change the sign and significance of its effect.
In the Appendix, we report the result for a fifth model where we keep only \texttt{CORE} and remove all the other centrality measures.
Also, in this model the significance and sign of \texttt{CORE} do not change (see Table~\ref{tab:model5}).

Finally, we note that \textit{Model 3} and \textit{Model 4} have about the same explanatory power (Log-Likelihood), but both are better than \textit{Model 2}, i.e., the model without coreness.
To make test this statement more rigorously, we use the Vuong closeness test~\citep{vuong1989likelihood}.
This test allows us to compare different models, perform model selection, and tell us which model captures the data better.
In Table~\ref{tab:vuong}, we report the results of the Vuong closeness test and find that \textit{Model 3} and \textit{Model 4} are both significantly better than \textit{Model 2}.
However, we cannot decided between \textit{Model 3} and \textit{Model 4}.
Overall, our analyses confirm that \texttt{CORE} is a good explanatory variable for firms' innovation output.

\begin{table}
  \footnotesize
  \begin{center}
  \begin{tabular}{l D{)}{)}{9)3} D{)}{)}{9)3} D{)}{)}{9)3} D{)}{)}{9)3}}
  \toprule
   & \multicolumn{1}{c}{Model 1} & \multicolumn{1}{c}{Model 2} & \multicolumn{1}{c}{Model 3} & \multicolumn{1}{c}{Model 4} \\
  \midrule
  Zero model: (Intercept)                                & 1.25 \; (0.07)^{***}  & 1.25 \; (0.07)^{***}  & 1.28 \; (0.21)^{***}  & 1.28 \; (0.21)^{***}  \\
  Zero model: log(PAT + 1)                               & -1.13 \; (0.04)^{***} & -1.13 \; (0.04)^{***} & -1.13 \; (0.07)^{***} & -1.14 \; (0.07)^{***} \\
  \midrule
  (Intercept)                               & -0.38 \; (0.12)^{**}  & -0.32 \; (0.12)^{**}  & -0.31 \; (0.12)^{**}  & -0.30 \; (0.11)^{**}  \\
  log(PAT + 1)                              & 0.90 \; (0.01)^{***}  & 0.89 \; (0.01)^{***}  & 0.88 \; (0.01)^{***}  & 0.87 \; (0.01)^{***}  \\
  EFF                                       & -0.02 \; (0.01)       & 0.02 \; (0.03)        & 0.05 \; (0.02)^{**}   &                       \\
  BETWEENNESS\_NORM                         & 0.02 \; (0.01)^{***}  & 0.01 \; (0.01)        & 0.01 \; (0.01)        & 0.00 \; (0.01)        \\
  DEGREE                                    &                       & 0.02 \; (0.01)        & -0.02 \; (0.01)       & -0.02 \; (0.01)       \\
  LOCAL\_REACH                              &                       & 0.05 \; (0.02)^{**}   & 0.03 \; (0.02)        & 0.02 \; (0.02)        \\
  LOCAL\_CLUSTERING                         &                       & 0.03 \; (0.03)        & 0.05 \; (0.05)        & 0.01 \; (0.03)        \\
  CORE                                    &                       &                       & -0.10 \; (0.03)^{***} & -0.09 \; (0.02)^{***} \\
  \midrule
  AIC                                                    & 48419.62              & 48410.93              & 48344.97              & 48347.99              \\
  Log Likelihood                                         & -24180.81             & -24173.47             & -24139.48             & -24143.99             \\
  Num. obs.                                              & 12649                 & 12649                 & 12649                 & 12649                 \\
  \bottomrule
  \multicolumn{5}{l}{\scriptsize{$^{***}p<0.001$, $^{**}p<0.01$, $^*p<0.05$.
  The and year fixed effect are reported in Appendix.}}\\
  \multicolumn{5}{l}{\scriptsize{The errors have been clustered at CORE (21 classes).}}
  \end{tabular}
  \caption{Zero-inflated negative binomial models.
  For this regression, we have standardized the variable in order to be able to compare the effect sizes.
  For CORE, the average value is around $16.53$, and the standard deviation is around $2.82$.}\label{tab:models}
  \end{center}
  \end{table}

\begin{table}
  \centering
  \footnotesize
  \begin{tabular}{rcrcll}
  \toprule
  $\,$ & Vuong z-statistic & $\,$             & $H_\textrm{A}$ &     $\,$        & $p$-value \\
  \midrule
  Raw  &   -3.357278 & \textit{Model 3} & > & \textit{Model 2} & 0.00039357\\
  AIC-corrected    & -3.258480 & \textit{Model 3} & > & \textit{Model 2} & 0.00056005 \\
  BIC-corrected    &     -2.890686 & \textit{Model 3} & > & \textit{Model 2} & 0.00192201\\
  \midrule
  Raw  &   -3.24855 & \textit{Model 4} & > & \textit{Model 2} & 0.00057997\\
  AIC-corrected    & -3.24855 & \textit{Model 4} & > & \textit{Model 2} & 0.00057997 \\
  BIC-corrected    &     -3.24855 & \textit{Model 4} & > & \textit{Model 2} & 0.00057997\\
  \midrule
  Raw  &   1.1754242 & \textit{Model 3} & > & \textit{Model 4} & 0.11991 \\
  AIC-corrected    & 0.6598649 & \textit{Model 3} & > & \textit{Model 4} & 0.25467 \\
  BIC-corrected    &     -1.2593904 & \textit{Model 3} & < & \textit{Model 4} & 0.10394\\
  \bottomrule
  \end{tabular}
  \caption{Results of the Vuong closeness test.}\label{tab:vuong}
\end{table}

\section{Discussion}\label{sec:discussion}

\paragraph{Firms' embeddedness and its evolution.}
In this paper, we argue that the embeddedness of firms in the R\&D network is indicative of their ability to innovate.
The latter is proxied by the number of patents firms file in a given time window.
To proxy embeddedness, we introduce a new measure, coreness.
It is a relative measure to compare firms' positions in the network at different stages of the evolution.
To calculate coreness, we use the \emph{weighted $k$-core decomposition} \citep{Garas2012a}, which takes into account multiple R\&D alliances between firms.
The decomposition applies a sequence of node removals to prune the network and assigns a value to each firm that indicates its current distance from the core.
In this respect, it reflects the embeddedness of firms better than previously used network centrality measures.

As the R\&D network evolves both as new firms enter and new alliances are formed, firms' coreness values change, reflecting changes in their network positions.
We observe the emergence of a clear core-periphery structure characterized by a dense core containing a smaller number of firms and a sparse periphery containing the majority of less integrated firms.
Firms that have reached the core of the alliance network are shown to be more successful in their innovation output than firms in the periphery.
Analyzing the relation between the coreness values of firms and the number of patents, we find a strong correlation.

\paragraph{Embeddedness and innovation.}
To better quantify the higher innovation output coming from higher embeddedness, we have performed a regression.
We have used a zero-inflated negative binomial regression to model firms' innovation output, measured by the number of patents filed in the subsequent year.
We find that decreasing the coreness by one unit increases the logarithm of the patent count by $\sim0.09/2.82=0.03$ (see Table~\ref{tab:models}).
This effect size implies that, for example, given a firm with coreness $C$ filing 100 patents, an identical copy of this firm with coreness $C-1$ would file about 103 patents.
Another example: Given a firm with coreness $C$ filing 10 patents, an identical copy of this firm with coreness $C-1$ would file about 13.5 patents.
Our results are significant even after controlling for several factors, including firms' past innovation output~\citep{blundell1995dynamic}, ability to absorb knowledge flows in the alliance network~\citep{phelps2010alongitudinal}, the number of partners, network clustering and structural holes~\citep{ahuja2000collaboratio, baum2000don}.
Overall, our results suggest that embeddedness has a positive and significant effect on innovation output in R\&D activities.

\paragraph{Comparison with previous works.}
Similar to \citep{schilling2007interfirm}, our regression analysis uses negative binomial as firms' patent count is an over-dispersed variable.
Different from \citep{schilling2007interfirm}, we use a zero-inflated version of the negative binomial as 50\% of the firms in our sample file 0 patents in a year.
Similar to~\citep{powell1999network}, we find that embeddedness has a significant effect on firms innovation output.
However, different from~\citep{powell1999network}, we do not quantify firms' embeddedness using the Katz-Bonacich centrality but use a new indicator, \emph{coreness}.
It is based on the $k$-core decomposition as in~\citep{al2011chapter} but takes multiple relations between firms into account.
This decision was motivated by the fact that Katz-Bonacich centrality may fail to measure embeddedness (see Sect.~\ref{SEC:Methods}).

Following~\citep{tomasello2014therole,Tomasello2013,vaccario2018quantifying}, we consider an alliance network with firms across industrial sectors.
That means, in contrast to the results of~\citet{powell1999network,al2011chapter}, our findings are not restricted to the biotechnology sector.
Firms' innovation might depend on complementary capabilities coming from any industrial sector, and hence, firms' embeddedness should be quantified in a network containing all the industrial sectors.
By doing this, the presented results are more general.

\paragraph{Limitations and outlooks.}
The analysed data is limited since it contains only collaborations until 2009 and patent data until 2006.
At the same time, to our knowledge, we have performed the largest analysis to quantify the link between the firms' embeddedness and their innovation output, with more than 13,000 firms over a time horizon of 25 years.
We have performed various statistical analyses to ensure the robustness of our analysis.

Given the evolution of the alliance network \citep{Tomasello2013} and firms' embeddedness (see Sect.~\ref{sec:link-centr-succ}), further studies could investigate how quickly the innovation output of firms change after they have obtained a more central position.
To extend our setup, we could add a regression analysis also considering the patents filed after two and three years and investigate how the effect size changes.

{\small \setlength{\bibsep}{1pt}
\bibliography{GTS-RND_CareerPaths}

\begin{thebibliography}{}

\bibitem[\protect\citeauthoryear{Ahuja}{Ahuja}{2000}]{ahuja2000collaboratio}
Ahuja, G. (2000).
\newblock Collaboration networks, structural holes, and innovation: A
  longitudinal study.
\newblock {\em Administrative Science Quarterly\/}~{\em 45\/}(3), 425--455.

\bibitem[\protect\citeauthoryear{Al-Laham and Bort}{Al-Laham and
  Bort}{2011}]{al2011chapter}
Al-Laham, A. and S.~Bort (2011).
\newblock Chapter 13 the innovation outcomes of mnc subsidiaries' local
  embeddedness: Evidence from the german "bioregion rhein-neckar-dreieck'local
  network".
\newblock {\em Entrepreneurship in the Global Firm (Progress in International
  Business Research, Volume 6). Emerald Group Publishing Limited\/}, 291--323.

\bibitem[\protect\citeauthoryear{Bastian, Heymann, and Jacomy}{Bastian
  et~al.}{2009}]{gephi}
Bastian, M., S.~Heymann, and M.~Jacomy (2009).
\newblock Gephi: An open source software for exploring and manipulating
  networks.
\newblock In {\em International AAAI Conference on Weblogs and Social Media}.

\bibitem[\protect\citeauthoryear{Baum, Calabrese, and Silverman}{Baum
  et~al.}{2000}]{baum2000don}
Baum, J.~A., T.~Calabrese, and B.~S. Silverman (2000).
\newblock Don't go it alone: Alliance network composition and startups'
  performance in canadian biotechnology.
\newblock {\em Strategic management journal\/}~{\em 21\/}(3), 267--294.

\bibitem[\protect\citeauthoryear{Blundell, Griffith, and Reenen}{Blundell
  et~al.}{1995}]{blundell1995dynamic}
Blundell, R., R.~Griffith, and J.~V. Reenen (1995).
\newblock Dynamic count data models of technological innovation.
\newblock {\em The Economic Journal\/}~{\em 105\/}(429), 333--344.

\bibitem[\protect\citeauthoryear{Bollob{\'a}s}{Bollob{\'a}s}{1988}]{bollobas1989graph}
Bollob{\'a}s, B. (1988).
\newblock {\em Graph theory and combinatorics 1988}.
\newblock Elsevier.

\bibitem[\protect\citeauthoryear{Borgatti and Everett}{Borgatti and
  Everett}{2000}]{Borgatti2000}
Borgatti, S.~P. and M.~G. Everett (2000).
\newblock Models of core/periphery structures.
\newblock {\em Social Networks\/}~{\em 21\/}(4), 375 -- 395.

\bibitem[\protect\citeauthoryear{Burt}{Burt}{2009}]{burt2009structural}
Burt, R.~S. (2009).
\newblock {\em Structural holes: The social structure of competition}.
\newblock Harvard university press.

\bibitem[\protect\citeauthoryear{Carmi, Havlin, Kirkpatrick, Shavitt, and
  Shir}{Carmi et~al.}{2007}]{Carmi2007}
Carmi, S., S.~Havlin, S.~Kirkpatrick, Y.~Shavitt, and E.~Shir (2007, July).
\newblock {A model of Internet topology using k-shell decomposition.}
\newblock {\em Proceedings of the National Academy of Sciences of the United
  States of America\/}~{\em 104\/}(27), 11150--4.

\bibitem[\protect\citeauthoryear{Freeman}{Freeman}{1991}]{freeman1991networks}
Freeman, C. (1991).
\newblock Networks of innovators: a synthesis of research issues.
\newblock {\em Research policy\/}~{\em 20\/}(5), 499--514.

\bibitem[\protect\citeauthoryear{Freeman}{Freeman}{1979}]{Freeman1979}
Freeman, L.~C. (1979).
\newblock {Centrality in social networks conceptual clarification}.
\newblock {\em Social Networks\/}~{\em 1\/}(3), 215--239.

\bibitem[\protect\citeauthoryear{Garas, Argyrakis, Rozenblat, Tomassini, and
  Havlin}{Garas et~al.}{2010}]{Garas2010}
Garas, A., P.~Argyrakis, C.~Rozenblat, M.~Tomassini, and S.~Havlin (2010,
  November).
\newblock {Worldwide spreading of economic crisis}.
\newblock {\em New Journal of Physics\/}~{\em 12\/}(11), 113043.

\bibitem[\protect\citeauthoryear{Garas, Schweitzer, and Havlin}{Garas
  et~al.}{2012}]{Garas2012a}
Garas, A., F.~Schweitzer, and S.~Havlin (2012, August).
\newblock {A $k$-shell decomposition method for weighted networks}.
\newblock {\em New Journal of Physics\/}~{\em 14\/}(8), 083030.

\bibitem[\protect\citeauthoryear{Gilsing, Nooteboom, Vanhaverbeke, Duysters,
  and {van den Oord}}{Gilsing et~al.}{2008}]{gilsing2008network}
Gilsing, V., B.~Nooteboom, W.~Vanhaverbeke, G.~Duysters, and A.~{van den Oord}
  (2008).
\newblock Network embeddedness and the exploration of novel technologies:
  Technological distance, betweenness centrality and density.
\newblock {\em Research Policy\/}~{\em 37\/}(10), 1717--1731.
\newblock Special Section Knowledge Dynamics out of Balance: Knowledge Biased,
  Skewed and Unmatched.

\bibitem[\protect\citeauthoryear{Granovetter}{Granovetter}{1985}]{granovetter1985economic}
Granovetter, M. (1985).
\newblock Economic action and social structure: The problem of embeddedness.
\newblock {\em American journal of sociology\/}~{\em 91\/}(3), 481--510.

\bibitem[\protect\citeauthoryear{Gulati, Nohria, and Zaheer}{Gulati
  et~al.}{2000}]{gulati2000strategic}
Gulati, R., N.~Nohria, and A.~Zaheer (2000).
\newblock Strategic networks.
\newblock {\em Strategic management journal\/}~{\em 21\/}(3), 203--215.

\bibitem[\protect\citeauthoryear{Hall, Jaffe, and Trajtenberg}{Hall
  et~al.}{2001}]{hall2001nber}
Hall, B.~H., A.~B. Jaffe, and M.~Trajtenberg (2001).
\newblock The nber patent citation data file: Lessons, insights and
  methodological tools.

\bibitem[\protect\citeauthoryear{Kitsak, Gallos, Havlin, Liljeros, Muchnik,
  Stanley, and Makse}{Kitsak et~al.}{2010}]{Kitsak2010}
Kitsak, M., L.~K. Gallos, S.~Havlin, F.~Liljeros, L.~Muchnik, H.~E. Stanley,
  and H.~A. Makse (2010, August).
\newblock {Identification of influential spreaders in complex networks}.
\newblock {\em Nature Physics\/}~{\em 6\/}(11), 888--893.

\bibitem[\protect\citeauthoryear{Newman}{Newman}{2018}]{newman2018networks}
Newman, M. (2018).
\newblock {\em Networks}.
\newblock Oxford university press.

\bibitem[\protect\citeauthoryear{Newman}{Newman}{2010}]{newman2010introduction}
Newman, M. E.~J. (2010).
\newblock {\em Networks: an introduction}.
\newblock Oxford; New York: Oxford University Press.

\bibitem[\protect\citeauthoryear{Nooteboom}{Nooteboom}{1999}]{nooteboom1999inter}
Nooteboom, B. (1999).
\newblock {\em Inter-firm alliances: Analysis and design}.
\newblock Psychology Press.

\bibitem[\protect\citeauthoryear{Owen-Smith and Powell}{Owen-Smith and
  Powell}{2004}]{owen2004knowledge}
Owen-Smith, J. and W.~W. Powell (2004).
\newblock Knowledge networks as channels and conduits: The effects of
  spillovers in the boston biotechnology community.
\newblock {\em Organization science\/}~{\em 15\/}(1), 5--21.

\bibitem[\protect\citeauthoryear{Paier and Scherngell}{Paier and
  Scherngell}{2011}]{paier2011determinants}
Paier, M. and T.~Scherngell (2011).
\newblock Determinants of collaboration in european r\&d networks: Empirical
  evidence from a discrete choice model.
\newblock {\em Industry and Innovation\/}~{\em 18\/}(1), 89--104.

\bibitem[\protect\citeauthoryear{Phelps}{Phelps}{2010}]{phelps2010alongitudinal}
Phelps, C.~C. (2010).
\newblock A longitudinal study of the influence of alliance network structure
  and composition on firm exploratory innovation.
\newblock {\em Academy of Management Journal\/}~{\em 53\/}(4), 890--913.

\bibitem[\protect\citeauthoryear{Polanyi and MacIver}{Polanyi and
  MacIver}{1944}]{polanyi1944great}
Polanyi, K. and R.~M. MacIver (1944).
\newblock {\em The great transformation}, Volume~2.
\newblock Beacon press Boston.

\bibitem[\protect\citeauthoryear{Powell, Koput, Smith-Doerr, and
  Owen-Smith}{Powell et~al.}{1999}]{powell1999network}
Powell, W., K.~Koput, L.~Smith-Doerr, and J.~Owen-Smith (1999, 01).
\newblock Network position and firm performance: Organizational returns to
  collaboration in the biotechnology industry.
\newblock {\em Research in the Sociology of Organizations\/}~{\em 16}.

\bibitem[\protect\citeauthoryear{Powell and Brantley}{Powell and
  Brantley}{1992}]{nohria1992networks}
Powell, W.~W. and P.~Brantley (1992).
\newblock {\em Networks and Organizations: Structure, Form, and Action (Chapter
  14)}.
\newblock Harvard Business School Press.

\bibitem[\protect\citeauthoryear{Schilling and Phelps}{Schilling and
  Phelps}{2007}]{schilling2007interfirm}
Schilling, M.~A. and C.~C. Phelps (2007).
\newblock Interfirm collaboration networks: The impact of large-scale network
  structure on firm innovation.
\newblock {\em Management science\/}~{\em 53\/}(7), 1113--1126.

\bibitem[\protect\citeauthoryear{Schweitzer, Garas, Tomasello, Vaccario, and
  Verginer}{Schweitzer et~al.}{2021}]{schweitzer-garas-2021}
Schweitzer, F., A.~Garas, M.~V. Tomasello, G.~Vaccario, and L.~Verginer (2021).
\newblock The role of network embeddedness on the selection of collaboration
  partners: An agent-based model with empirical validation.
\newblock {\em Advances in Complex Systems\/}~{\em 24\/}(7-8), xxxx.

\bibitem[\protect\citeauthoryear{Seidman}{Seidman}{1983}]{seidman1983network}
Seidman, S.~B. (1983).
\newblock Network structure and minimum degree.
\newblock {\em Social Networks\/}~{\em 5\/}(3), 269 -- 287.

\bibitem[\protect\citeauthoryear{Shan, Walker, and Kogut}{Shan
  et~al.}{1994}]{shan1994interfirm}
Shan, W., G.~Walker, and B.~Kogut (1994).
\newblock Interfirm cooperation and startup innovation in the biotechnology
  industry.
\newblock {\em Strategic Management Journal\/}~{\em 15\/}(5), 387--394.

\bibitem[\protect\citeauthoryear{Thomson-Reuters}{Thomson-Reuters}{2013}]{sdc2013}
Thomson-Reuters (2013).
\newblock Sdc platinum dataset.
\newblock Date of access: 07/04/2014.

\bibitem[\protect\citeauthoryear{Tomasello, Napoletano, Garas, and
  Schweitzer}{Tomasello et~al.}{2016}]{Tomasello2013}
Tomasello, M.~V., M.~Napoletano, A.~Garas, and F.~Schweitzer (2016).
\newblock {The Rise and Fall of R\&D Networks}.
\newblock {\em Industrial and Corporate Change\/}~{\em dtw041}.

\bibitem[\protect\citeauthoryear{Tomasello, Perra, Tessone, Karsai, and
  Schweitzer}{Tomasello et~al.}{2014}]{tomasello2014therole}
Tomasello, M.~V., N.~Perra, C.~J. Tessone, M.~Karsai, and F.~Schweitzer (2014).
\newblock The role of endogenous and exogenous mechanisms in the formation of
  r\&d networks.
\newblock {\em Scientific Reports\/}~{\em 4}.

\bibitem[\protect\citeauthoryear{Uzzi}{Uzzi}{1997}]{uzzi1997social}
Uzzi, B. (1997).
\newblock Social structure and competition in interfirm networks: The paradox
  of embeddedness.
\newblock {\em Administrative science quarterly\/}, 35--67.

\bibitem[\protect\citeauthoryear{Vaccario, Tomasello, Tessone, and
  Schweitzer}{Vaccario et~al.}{2018}]{vaccario2018quantifying}
Vaccario, G., M.~V. Tomasello, C.~J. Tessone, and F.~Schweitzer (2018, August).
\newblock Quantifying knowledge exchange in r\&d networks: A data-driven model.
\newblock {\em Journal of Evolutionary Economics\/}~{\em 28\/}(3), 461--493.

\bibitem[\protect\citeauthoryear{Vuong}{Vuong}{1989}]{vuong1989likelihood}
Vuong, Q.~H. (1989).
\newblock Likelihood ratio tests for model selection and non-nested hypotheses.
\newblock {\em Econometrica\/}~{\em 57\/}(2), 307--333.

\end{thebibliography}
\bibliographystyle{chicago}
}

\clearpage

\section*{Appendix}

\begin{appendix}

\section{Industry and time effects.}

In Table~\ref{tab:models_full} we report all the control variable of \textit{Model 1}, \textit{Model 2}, \textit{Model 3}, and \textit{Model 4}.
\begin{table}
  \footnotesize
  \begin{center}
  \begin{tabular}{l D{)}{)}{9)3} D{)}{)}{9)3} D{)}{)}{9)3} D{)}{)}{9)3}}
  \toprule
   & \multicolumn{1}{c}{\textit{Model 1}} & \multicolumn{1}{c}{\textit{Model 2}} & \multicolumn{1}{c}{\textit{Model 3}} & \multicolumn{1}{c}{\textit{Model 4}} \\
  \midrule
   (Intercept)                               & -0.38 \; (0.12)^{**}  & -0.32 \; (0.12)^{**}  & -0.35 \; (0.16)^{*}   & -0.35 \; (0.16)^{*}   \\
   log(PAT + 1)                              & 0.90 \; (0.01)^{***}  & 0.89 \; (0.01)^{***}  & 0.87 \; (0.02)^{***}  & 0.87 \; (0.02)^{***}  \\
    I\_automotive bodies and parts       & -0.21 \; (0.06)^{***} & -0.18 \; (0.06)^{**}  & -0.16 \; (0.12)       & -0.17 \; (0.12)       \\
    I\_chemicals                         & -0.60 \; (0.05)^{***} & -0.57 \; (0.05)^{***} & -0.50 \; (0.03)^{***} & -0.50 \; (0.03)^{***} \\
    I\_computer and office equipment     & -0.29 \; (0.05)^{***} & -0.29 \; (0.05)^{***} & -0.34 \; (0.07)^{***} & -0.34 \; (0.07)^{***} \\
    I\_household audiovisual equipment   & -0.12 \; (0.07)       & -0.10 \; (0.07)       & -0.19 \; (0.13)       & -0.20 \; (0.13)       \\
    I\_measuring and controlling devices & -0.46 \; (0.06)^{***} & -0.44 \; (0.06)^{***} & -0.40 \; (0.05)^{***} & -0.40 \; (0.06)^{***} \\
    I\_medical equipment                 & -0.42 \; (0.05)^{***} & -0.39 \; (0.05)^{***} & -0.34 \; (0.06)^{***} & -0.34 \; (0.06)^{***} \\
    I\_petroleum refining and products   & -0.80 \; (0.12)^{***} & -0.76 \; (0.12)^{***} & -0.72 \; (0.13)^{***} & -0.72 \; (0.13)^{***} \\
    I\_pharmaceuticals                   & -0.40 \; (0.04)^{***} & -0.39 \; (0.04)^{***} & -0.39 \; (0.07)^{***} & -0.38 \; (0.07)^{***} \\
    I\_aerospace equipment               & -0.24 \; (0.07)^{***} & -0.22 \; (0.07)^{***} & -0.22 \; (0.14)       & -0.22 \; (0.14)       \\
    I\_telecommunications equipment      & -0.33 \; (0.05)^{***} & -0.33 \; (0.05)^{***} & -0.31 \; (0.08)^{***} & -0.30 \; (0.08)^{***} \\
   1988                                  & -0.03 \; (0.14)       & -0.03 \; (0.14)       & -0.08 \; (0.09)       & -0.07 \; (0.09)       \\
   1989                                  & -0.10 \; (0.14)       & -0.09 \; (0.14)       & -0.14 \; (0.12)       & -0.14 \; (0.12)       \\
   1990                                  & -0.15 \; (0.13)       & -0.14 \; (0.13)       & -0.19 \; (0.10)^{*}   & -0.19 \; (0.10)^{*}   \\
   1991                                  & -0.12 \; (0.12)       & -0.12 \; (0.12)       & -0.14 \; (0.12)       & -0.14 \; (0.12)       \\
   1992                                  & -0.15 \; (0.12)       & -0.18 \; (0.12)       & -0.16 \; (0.13)       & -0.16 \; (0.13)       \\
   1993                                  & -0.00 \; (0.12)       & -0.06 \; (0.12)       & -0.01 \; (0.12)       & -0.01 \; (0.12)       \\
   1994                                  & 0.17 \; (0.12)        & 0.08 \; (0.12)        & 0.16 \; (0.16)        & 0.16 \; (0.17)        \\
   1995                                  & -0.22 \; (0.12)       & -0.33 \; (0.12)^{**}  & -0.23 \; (0.12)       & -0.23 \; (0.12)^{*}   \\
   1996                                  & -0.15 \; (0.12)       & -0.25 \; (0.12)^{*}   & -0.16 \; (0.12)       & -0.15 \; (0.12)       \\
   1997                                  & -0.34 \; (0.12)^{**}  & -0.40 \; (0.12)^{***} & -0.31 \; (0.12)^{**}  & -0.32 \; (0.12)^{**}  \\
   1998                                  & -0.32 \; (0.12)^{**}  & -0.37 \; (0.12)^{**}  & -0.29 \; (0.10)^{**}  & -0.29 \; (0.10)^{**}  \\
   1999                                  & -0.40 \; (0.12)^{***} & -0.42 \; (0.12)^{***} & -0.36 \; (0.09)^{***} & -0.36 \; (0.09)^{***} \\
   EFF                                       & -0.02 \; (0.01)       & 0.02 \; (0.03)        & 0.05 \; (0.02)^{**}   &                       \\
   BETWEENNESS\_NORM                         & 0.02 \; (0.01)^{***}  & 0.01 \; (0.01)        & 0.01 \; (0.01)        & 0.01 \; (0.01)        \\
   Log(theta)                                & 0.56 \; (0.03)^{***}  & 0.57 \; (0.03)^{***}  & 0.58 \; (0.03)^{***}  & 0.58 \; (0.03)^{***}  \\
  Zero model: (Intercept)                                & 1.25 \; (0.07)^{***}  & 1.25 \; (0.07)^{***}  & 1.28 \; (0.21)^{***}  & 1.28 \; (0.21)^{***}  \\
  Zero model: log(PAT + 1)                               & -1.13 \; (0.04)^{***} & -1.13 \; (0.04)^{***} & -1.13 \; (0.07)^{***} & -1.13 \; (0.07)^{***} \\
   DEGREE                                    &                       & 0.02 \; (0.01)        & -0.02 \; (0.01)       & -0.02 \; (0.01)       \\
   LOCAL\_REACH                              &                       & 0.05 \; (0.02)^{**}   & 0.03 \; (0.02)        & 0.02 \; (0.02)        \\
   LOCAL\_CLUSTERING                         &                       & 0.03 \; (0.03)        & 0.05 \; (0.05)        & 0.01 \; (0.03)        \\
   CORE                                    &                       &                       & -0.10 \; (0.03)^{***} & -0.09 \; (0.02)^{***} \\
  \midrule
  AIC                                                    & 48419.62              & 48410.93              & 48344.97              & 48347.53              \\
  Log Likelihood                                         & -24180.81             & -24173.47             & -24139.48             & -24141.76             \\
  Num. obs.                                              & 12649                 & 12649                 & 12649                 & 12649                 \\
  \bottomrule
  \multicolumn{5}{l}{\scriptsize{$^{***}p<0.001$; $^{**}p<0.01$; $^{*}p<0.05$}}\\
  \multicolumn{5}{l}{\scriptsize{The errors have been clustered at CORE (21 classes).}}
  \end{tabular}
  \caption{Zero-inflated negative binomial models.}
  \label{tab:models_full}
  \end{center}
  \end{table}

  In Table~\ref{tab:model5} we report all the control variable of \textit{Model 3}, \textit{Model 4}, and \textit{Model 5}.
  \begin{table}
    \footnotesize
    \begin{center}
    \begin{tabular}{l D{)}{)}{9)3} D{)}{)}{9)3} D{)}{)}{9)3}}
    \toprule
     & \multicolumn{1}{c}{\textit{Model 3}} & \multicolumn{1}{c}{\textit{Model 4}} & \multicolumn{1}{c}{\textit{Model 5}} \\
    \midrule
    (Intercept)                               & -0.35 \; (0.16)^{*}   & -0.35 \; (0.16)^{*}   & -0.35 \; (0.16)^{*}   \\
    log(PAT + 1)                              & 0.87 \; (0.02)^{***}  & 0.87 \; (0.02)^{***}  & 0.87 \; (0.02)^{***}  \\
    I\_automotive bodies and parts       & -0.16 \; (0.12)       & -0.17 \; (0.12)       & -0.17 \; (0.12)       \\
    I\_chemicals                         & -0.50 \; (0.03)^{***} & -0.50 \; (0.03)^{***} & -0.51 \; (0.03)^{***} \\
    I\_computer and office equipment     & -0.34 \; (0.07)^{***} & -0.34 \; (0.07)^{***} & -0.34 \; (0.07)^{***} \\
    I\_household audiovisual equipment   & -0.19 \; (0.13)       & -0.20 \; (0.13)       & -0.20 \; (0.12)       \\
    I\_measuring and controlling devices & -0.40 \; (0.05)^{***} & -0.40 \; (0.06)^{***} & -0.41 \; (0.05)^{***} \\
    I\_medical equipment                 & -0.34 \; (0.06)^{***} & -0.34 \; (0.06)^{***} & -0.35 \; (0.05)^{***} \\
    I\_petroleum refining and products   & -0.72 \; (0.13)^{***} & -0.72 \; (0.13)^{***} & -0.74 \; (0.14)^{***} \\
    I\_pharmaceuticals                   & -0.39 \; (0.07)^{***} & -0.38 \; (0.07)^{***} & -0.38 \; (0.07)^{***} \\
    I\_aerospace equipment               & -0.22 \; (0.14)       & -0.22 \; (0.14)       & -0.22 \; (0.14)       \\
    I\_telecommunications equipment      & -0.31 \; (0.08)^{***} & -0.30 \; (0.08)^{***} & -0.31 \; (0.08)^{***} \\
    1988                                  & -0.08 \; (0.09)       & -0.07 \; (0.09)       & -0.07 \; (0.09)       \\
    1989                                  & -0.14 \; (0.12)       & -0.14 \; (0.12)       & -0.14 \; (0.13)       \\
    1990                                  & -0.19 \; (0.10)^{*}   & -0.19 \; (0.10)^{*}   & -0.19 \; (0.10)       \\
    1991                                  & -0.14 \; (0.12)       & -0.14 \; (0.12)       & -0.14 \; (0.12)       \\
    1992                                  & -0.16 \; (0.13)       & -0.16 \; (0.13)       & -0.16 \; (0.13)       \\
    1993                                  & -0.01 \; (0.12)       & -0.01 \; (0.12)       & 0.00 \; (0.12)        \\
    1994                                  & 0.16 \; (0.16)        & 0.16 \; (0.17)        & 0.17 \; (0.17)        \\
    1995                                  & -0.23 \; (0.12)       & -0.23 \; (0.12)^{*}   & -0.20 \; (0.12)       \\
    1996                                  & -0.16 \; (0.12)       & -0.15 \; (0.12)       & -0.13 \; (0.12)       \\
    1997                                  & -0.31 \; (0.12)^{**}  & -0.32 \; (0.12)^{**}  & -0.30 \; (0.12)^{**}  \\
    1998                                  & -0.29 \; (0.10)^{**}  & -0.29 \; (0.10)^{**}  & -0.29 \; (0.10)^{**}  \\
    1999                                  & -0.36 \; (0.09)^{***} & -0.36 \; (0.09)^{***} & -0.36 \; (0.09)^{***} \\
    EFF                                       & 0.05 \; (0.02)^{**}   &                       &                       \\
    BETWEENNESS\_NORM                         & 0.01 \; (0.01)        & 0.01 \; (0.01)        &                       \\
    DEGREE                                    & -0.02 \; (0.01)       & -0.02 \; (0.01)       &                       \\
    LOCAL\_REACH                              & 0.03 \; (0.02)        & 0.02 \; (0.02)        &                       \\
    LOCAL\_CLUSTERING                         & 0.05 \; (0.05)        & 0.01 \; (0.03)        &                       \\
    CORE                                    & -0.10 \; (0.03)^{***} & -0.09 \; (0.02)^{***} & -0.09 \; (0.02)^{***} \\
    Log(theta)                                & 0.58 \; (0.03)^{***}  & 0.58 \; (0.03)^{***}  & 0.58 \; (0.03)^{***}  \\
    Zero model: (Intercept)                                & 1.28 \; (0.21)^{***}  & 1.28 \; (0.21)^{***}  & 1.28 \; (0.21)^{***}  \\
    Zero model: log(PAT + 1)                               & -1.13 \; (0.07)^{***} & -1.13 \; (0.07)^{***} & -1.14 \; (0.07)^{***} \\
    \midrule
    AIC                                                    & 48344.97              & 48347.53              & 48345.42              \\
    Log Likelihood                                         & -24139.48             & -24141.76             & -24144.71             \\
    Num. obs.                                              & 12649                 & 12649                 & 12649                 \\
    \bottomrule
    \multicolumn{4}{l}{\scriptsize{$^{***}p<0.001$; $^{**}p<0.01$; $^{*}p<0.05$}}\\
    \multicolumn{4}{l}{\scriptsize{The errors have been clustered at CORE (21 classes).}}
    \end{tabular}
    \caption{Zero-inflated negative binomial models.}\label{tab:model5}
    \end{center}
    \end{table}
\end{appendix}

\end{document}